\documentclass[aps,prb,reprint,superscriptaddress]{revtex4-2}

\usepackage{gensymb}
\usepackage{placeins}
\usepackage{xcolor}

\usepackage{amsmath}
\usepackage{mathtools}
\usepackage{amsfonts}
\usepackage{amssymb}
\usepackage{relsize}
\usepackage{graphicx}
\usepackage{makecell}
\usepackage{float}
\usepackage{tabularx}
\usepackage{multirow}

\newcommand\lfrac[2]{\frac{\displaystyle #1}{\displaystyle #2}}

\begin{document}

\title{Monte Carlo Spin Simulations of Magnetic Noise - The Search for Pivoting}
\author{D. L. Mickelsen}
\affiliation{Department of Physics and Astronomy, University of California, Irvine, California, 92697}
\author{Ruqian Wu}
\affiliation{Department of Physics and Astronomy, University of California, Irvine, California, 92697}
\author{Clare C. Yu}
\affiliation{Department of Physics and Astronomy, University of California, Irvine, California, 92697}

\date{\today}

\begin{abstract}
Superconducting quantum interference devices (SQUIDs) show great promise as quantum bits (qubits) but continue to be hindered by flux noise.  The flux noise power spectra of SQUIDs go as $1/f^\alpha$, where $\alpha$ is the temperature-dependent noise exponent.  Experiments find $0.5 \lesssim \alpha \lesssim 1$.  Furthermore, experiments find that the noise power spectra versus frequency at different temperatures pivot about or cross at a common point for each SQUID.  To try to better understand the results and motivated by experimental evidence that magnetic moments on the surface of SQUIDS produce flux noise, we present the results of our Monte Carlo simulations of various spin systems on 2D lattices.  We find that only spin glasses produce $\alpha \sim 1$ at low temperature.  We find that aliasing of the noise power spectra at high frequencies can lead to spectral pivoting if it is in proximity to a knee at a slightly lower frequency.  We show that the pivot frequency depends on the method of site selection and how often the magnetization is recorded.  The spectral pivoting that occurs in our simulations is due to aliasing and does not explain the spectral pivoting of experiments.
\end{abstract}

\newcommand{\beq}{\begin{eqnarray}}
\newcommand{\eeq}{\end{eqnarray}}
\maketitle

\section{Introduction}

Superconducting quantum interference devices (SQUIDs) can be used as quantum bits (qubits).  While SQUIDs hold great potential for quantum computing, they suffer from noise and decoherence.  One of the main sources of decoherence is $1/f$ flux noise~\cite{Bialczak2007}.  $1/f$ noise is characterized by noise power spectra that go as $1/f^\alpha$, where $f$ is frequency and $\alpha$ is the noise exponent.  The noise exponents for SQUIDs lie in the range $0.5 \lesssim \alpha \lesssim 1$ for $1\text{ K} \gtrsim T \gtrsim 20 \text{ mK}$~\cite{Wellstood1987,Anton2013,Kempf2016}, where the noise exponent increases as temperature decreases.  Fluctuating magnetic moments were proposed as a source of flux noise in SQUIDs~\cite{Wellstood1987}.  This is consistent with experimental evidence of surface spins on normal metals~\cite{Bluhm2009} and superconductors~\cite{Sendelbach2008}.  Sendelbach {\it et al.} measured a $1/T$ temperature-dependent flux through SQUIDs which is indicative of paramagnetic spins~\cite{Sendelbach2008}.  Fluctuations of these surface spins cause flux noise because SQUIDs are highly sensitive magnetometers. Furthermore, cross correlations between fluctuations in the flux and the inductance in DC SQUIDs indicate that the spins interact ferromagnetically \cite{Sendelbach2009}.

Additional evidence for surface spins comes from density functional theory (DFT) calculations of the oxide layer on the aluminum surface of SQUIDs.  Since SQUIDs are exposed to air, it is reasonable to expect oxygen to be adsorbed on the surface below approximately $40 \text{ K}$.  $\text{O}_2$ is paramagnetic, and DFT calculations show that oxygen retains a magnetic moment of $1.8\text{ }\mu_B$ after being adsorbed on the surface of sapphire ($\alpha$-Al\textsubscript{2}O\textsubscript{3})~\cite{Wang2015}.  DFT also finds a low barrier ($\sim 10 \text{ mK}$) for spin reorientation so that spins rotate easily in the easy plane that is perpendicular to the $\text{O}_2$ molecular bond~\cite{Wang2015,Kumar2016}.  In addition, these calculations indicate that the coupling between adsorbed oxygen molecules is ferromagnetic~\cite{Wang2015}.  Monte Carlo simulations of ferromagnetically coupled $\text{O}_2$ on the surface of sapphire are able to produce $1/f$ noise consistent with experiment at higher temperatures~\cite{Wang2015}.  Evidence of paramagnetic oxygen spins on SQUIDs come from X-ray absorption spectroscopy (XAS) and x-ray magnetic circular dichroism (XMCD) experiments that were carried out on thin films of aluminum and niobium which are typical SQUID materials~\cite{Kumar2016}.  The experiments~\cite{Kumar2016} confirm the DFT predictions~\cite{Wang2015} that the bond axis of $\text{O}_2$ is tilted $55^\circ$ away from the surface normal.

Surface treatments that remove or prevent oxygen adsorption on SQUIDs reduce flux noise by a factor of four or five~\cite{Kumar2016}.  A protective coating of nonmagnetic ammonia ($\text{NH}_3$) prevents the adsorption of $\text{O}_2$ since $\text{NH}_3$ has a higher binding energy to $\text{Al}_2\text{O}_3$.  Ultraviolet irradiation of SQUIDs in an ultrahigh vacuum removes adsorbed oxygen~\cite{Kumar2016}.  Although flux noise is reduced with these treatments, it is not eliminated.

There is still the question as to how surface spins produce $1/f$ noise.  The Dutta-Horn model of $1/f$ noise~\cite{Dutta1981} is the most common explanation for $1/f$-type noise.  The model assumes that independent, thermally-activated processes exist, where each process individually produces a Lorentzian power spectra.  A distribution of barrier heights that is slowly varying on the order of $k_B T$ leads to a range of relaxation times and hence, $1/f$ noise.  In the case of spins, spin flips can result from spins hopping in and out of traps with different preferential spin orientations \cite{Koch2007} or hopping between a local moment state and the conduction band at the interface between a metal and an insulator \cite{Choi2009}. De Sousa proposed that magnetic noise arises from spin flips of paramagnetic dangling bonds at the amorphous semiconductor/oxide interface \cite{deSousa2007}. A range of relaxation times can also come from interactions between spins \cite{Faoro2008,Chen2010,De2014,De2019}. Faoro and Ioffe invoked spin diffusion via RKKY interactions between spins. Chen and Yu suggested that interacting surface spins can be modeled as a spin glass \cite{Chen2010} while De proposed a model with small spin clusters in which the spins have ferromagnetic RKKY interactions \cite{De2014,De2019}. These theoretical models of spins on the surface of SQUIDs have indeed been able to produce $1/f$ flux noise~\cite{Koch2007,deSousa2007,Faoro2008,Choi2009,Chen2010,De2014,De2019}.

While it is well established that surface spins are a source of $1/f$ flux noise, there are still mysteries associated with this noise.  For example, why is the noise exponent close to 1?  While Ising spin glass models yield $\alpha \lesssim 1$~\cite{Chen2010}, ferromagnetic spin system simulations find $\alpha > 1$ at low temperatures~\cite{Wang2015}.

Another puzzle comes from experiments using Nb SQUIDs. In this case, when the noise power spectra are plotted as a function of frequency on a log-log plot, the curves for different temperatures cross or intersect in the vicinity of a single ``pivot" frequency~\cite{Anton2013}. As a result, as the temperature decreases, the low frequency noise increases and the high frequency noise decreases. The noise near the pivot frequency is rather constant as a function of temperature. For SQUIDs with different geometries, the power spectra for each device still pivot.  The pivot frequency is different for each SQUID, and there is no clear relation between geometry and pivot frequency.  Other experiments involving SQUIDs of various materials also find spectral pivoting~\cite{Kempf2016}.

Futhermore,
Anton {\it et al.} \cite{Anton2013} integrated the (extrapolated) flux noise 
power spectra $S_{\Phi}(f)$ over the frequency $f$ from $f_1=10^{-4}$ Hz to
$f_2=10^9$ Hz to obtain the mean square flux noise $\langle\Phi^2\rangle$.
They found that the mean square flux noise increased by two to three orders of 
magnitude as the temperature increased from 0.1 K to 4 K.

Several models have been proposed to explain spectral pivoting.  Spin diffusion can explain pivoting if the system is close to a phase transition where the diffusion coefficient is dependent on temperature~\cite{Lanting2014}.  The result is a range of frequencies where power spectra cross as in experiment.  Monte Carlo simulations of Heisenberg spins in a cluster model can also produce spectral pivoting~\cite{Davis2018}.  This model bases the probability of spin flips on changes in free energy instead of internal energy as in the standard Monte Carlo method.  This has the effect that lower entropy spin configurations are more favored.  This model produces $1/f^\alpha$ power spectra at low frequency and $1/f^2$ power spectra at high frequency.  The power spectra do not intrinsically pivot.  To get crossing, the $1/f^\alpha$ parts of the power spectra are extended via extrapolation into the high-frequency range.

In an effort to understand spectral pivoting and what types of interactions lead to the noise exponent $\alpha \sim 1$, we performed Monte Carlo simulations of spins on 2D lattices since both DFT simulations~\cite{Wang2015} and experiments~\cite{Kumar2016} indicate spins can produce flux noise.  Since oxygen spins have an easy-plane anisotropy perpendicular to the $\text{O}_2$ bond~\cite{Wang2015}, we test various interacting spin models with different anisotropies.

Although we find that spectral pivoting can occur in Monte Carlo simulations of classical XY and Heisenberg spins, the pivoting in our simulations is an artifact of the simulations.  In plots of noise power versus frequency, a high frequency pivot occurs because the low-frequency knees are close in frequency to the high-frequency aliasing of the power spectra, and thus the simulations do not explain the experimentally observed pivoting. The low-frequency knee refers to the crossover from a flat noise spectrum at low frequencies to $1/f^{\alpha}$ at higher frequencies with the noise exponent $\alpha > 0$. 

Aliasing refers to a well-known phenomenon in signal processing. Suppose a signal is sampled at equal time intervals, $\Delta$. Then we can define a Nyquist frequency $f_c=1/(2\Delta)$. If we Fourier transform the time series, the result has components with frequencies outside the frequency range $-f_c<f<f_c$ that spuriously contribute to the Fourier transformed signal in the range between $-f_c$ and $f_c$ \cite{Press2007}. We say that the true power spectrum from frequencies outside that range are ``folded'' into the range. As a simple example, consider two waves: $g_1(t)=\exp(2\pi i f_1 t)$ and $g_2(t)=\exp(2\pi i f_2 t)$ where $f_2=f_1+(m/\Delta)$ where $m$ is a positive integer so that $f_2 > f_1$. It is easy to show that $g_1=g_2$ at time intervals of $\Delta$. So the higher frequency signal $g_2(t)$ will contribute to the Fourier transformed signal at lower frequencies such as $f_1$. In our simulations, aliasing appears as a minimum or flattening of the noise power spectra at $f_c$.


In addition to trying to better understand spectral pivoting,
we also use this paper to explain some of the challenges and pitfalls of Monte
Carlo simulations and what happens when there are deviations from 
the standard procedure.
The paper is structured in the following manner.  In Sec.~\ref{s_method}, we describe the Hamiltonians of the spin models, how the simulations are performed/equilibrated, and how the noise power is analyzed.  In Sec.~\ref{s_results}, we present our results on noise exponents, noise amplitudes, and why pivoting occurs in our simulations.

\section{Method}
\label{s_method}
\subsection{Spin Models}
The Ising, XY, and Heisenberg Hamiltonians of the 2D spin systems are given by
\begin{equation}
\label{e_Hamiltonian-Ising}
H_{\text{Ising}}=-\sum_{\langle i,j \rangle} J_{ij} s_i \cdot s_j
\end{equation}
\begin{equation}
\label{e_Hamiltonian-XY}
H_{\text{XY}}=-\sum_{\langle i,j \rangle} J_{ij} \mathbf{s}_i \cdot \mathbf{s}_j
\end{equation}
\begin{equation}
\label{e_Hamiltonian-Heisenberg}
H_{\text{Heis.}}=-\sum_{\langle i,j \rangle} J_{ij} \mathbf{s}_i \cdot \mathbf{s}_j-A \sum_i (\mathbf{n}_i \cdot \mathbf{s}_i)^2
\end{equation}
where $\mathbf{s}_i$ and $\mathbf{s}_j$ are classical spins of length 1 on nearest-neighbor sites $i$ and $j$, respectively.  $J_{ij}$ is the spin exchange coupling.  A positive value for $J_{ij}$ indicates a ferromagnetic interaction.  The second term in Eq.~(\ref{e_Hamiltonian-Heisenberg}) is a spin anisotropy term.  For each site $i$, there is a local anisotropy axis $\mathbf{n}_i$.  To model the disorder of the SQUID surface, $\mathbf{n}_i$ points in a direction that varies randomly from site to site.  The random-axis anisotropic model was proposed by Harris to describe magnetism in an amorphous material~\cite{Harris1973}.

If the anisotropy $A$ is positive, then $\mathbf{n}_i$ is the easy axis for spins; if the anisotropy $A$ is negative, then $\mathbf{n}_i$ is normal to the easy plane for spin orientation.

Six spin models were simulated: noninteracting spins ($J_{ij}=0$, $A=-1$), ferromagnet ($J_{ij}=1$), Poisson ferromagnet ($\langle J_{ij} \rangle=1$, $\sigma^2_{J_{ij}}=0.2$), antiferromagnet ($J_{ij}=-1$), spin glass ($\langle J_{ij} \rangle=0, \sigma_{J_{ij}}=1$), and spin glass ferromagnet ($\langle J_{ij} \rangle=0.5, \sigma_{J_{ij}}=1$).  While one would think that the problem of a noninteracting anisotropic Heisenberg spin could be solved exactly, this is not the case. However, one can make various approximations to the solution \cite{Brown1963}. For the Poisson ferromagnet, the couplings $J_{ij}$ are chosen in the following way ~\cite{Wang2015}. First random integers $C_{ij}$ are drawn from a Poisson distribution with a mean of 5.  Then the couplings are given by $J_{ij}$ = 0.2 $C_{ij}$ so that for the Poisson ferromagnet, $\langle J_{ij} \rangle = 1$ and $\sigma^2_{J_{ij}} = 0.2$.  The temperature for this system is measured in units of $\langle J_{ij} \rangle$.  

For the spin glass and spin glass ferromagnet, $J_{ij}$ is chosen from a Gaussian distribution with a variance $\sigma^2_{J_{ij}}$ and the temperature is measured in units of $\sigma_{J_{ij}}$. One way to obtain the random interactions associated with a spin glass is via RKKY interactions \cite{Faoro2008}. Spins trapped in local moment states at the interface between a metal and an insulator \cite{Choi2009} could interact via RKKY oscillations in the conduction electron gas of the metal. However, spins associated with adsorbates such as oxygen molecules on the surface of the native metal oxide are too far away from the conduction electrons of the metal to interact with each other via RKKY interactions \cite{Wang2015}. The interactions and the corresponding simulated spin models are summarized in Table~\ref{t_spin-models}.

\begin{table}
	\setlength{\tabcolsep}{6pt}
	\renewcommand{\arraystretch}{1.1}
	\centering
	\begin{tabular}{ |m{3.5cm}|m{3.1cm}| } 
		\hline
		Interaction & Spin Models \\
		\hline
		Noninteracting  & \multirow{2}{10em}{Heisenberg, $A=-1$} \\
		$J_{ij}=0$ & \\
		\hline
		\multirow{4}{9em}{Spin Glass $\langle J_{ij} \rangle=0$, $\sigma_{J_{ij}}=1$} & Heisenberg, $A=0$  \\ 
		& Heisenberg, $A=-10$  \\
		& XY  \\
		& Ising  \\
		\hline
		\multirow{3}{10em}{Antiferromagnet $J_{ij}=-1$} & Heisenberg, $A=0$  \\
		& Heisenberg, $A=-10$  \\
		& XY  \\
		\hline
		\multirow{3}{10em}{Poisson Ferromagnet $\langle J_{ij} \rangle=1$, $\sigma^2_{J_{ij}}=0.2$} & Heisenberg, $A=0$  \\
		& Heisenberg, $A=-10$  \\
		& XY  \\
		\hline
		\multirow{3}{8em}{Ferromagnet $J_{ij}=1$} & Heisenberg, $A=0$  \\
		& Heisenberg, $A=-10$  \\
		& XY  \\
		\hline
		\multirow{3}{15em}{Spin Glass Ferromagnet $\langle J_{ij} \rangle=0.5$, $\sigma_{J_{ij}}=1$} & Heisenberg, $A=0$  \\
		& Heisenberg, $A=-10$  \\
		& XY  \\
		\hline
	\end{tabular}
	\caption{All six types of spin exchange couplings $J_{ij}$ and the corresponding spins models that were simulated.}
	\label{t_spin-models}
\end{table}

\subsection{Simulation Details}
We perform simulations with spins occupying every site of a 16 x 16 square lattice.  The system is initialized with randomly oriented spins.  For anisotropic systems, each run has a unique, but random, set of anisotropic axes.  A spin is allowed to reorient itself according to the Metropolis algorithm~\cite{Metropolis1953}.  In this algorithm, a trial move consists of first choosing a spin on the lattice at random.  The initial energy of this site $E_i$ is calculated from the local field produced by its nearest neighbors and the local anisotropy.  A new orientation of the spin is chosen from a random distribution for the Heisenberg and XY systems.  For the Heisenberg systems, the distribution is random on the unit sphere (the distribution is uniform in $\phi$ and $\cos \theta$).  For XY systems, the distribution is random on the unit circle (uniform distribution in $\phi$).  In the case of Ising systems, the spin is flipped.  The final energy $E_f$ of this site with the new spin orientation is calculated.  If the final energy is less than the initial energy, then the new spin orientation is accepted.  However if the final energy is greater than the initial energy, then the flip is accepted with probability $\exp[-(E_f-E_i)/(k_B T)]$.  A random number is generated from a uniform distribution between 0 and 1; if it is less than the Boltzmann factor, then the new orientation is accepted.  This process continues for the remaining sites within the lattice.  The time it takes for one sweep through the lattice is one Monte Carlo step (MCS).

The system is equilibrated as described later in section~\ref{s_equilibration}. The total magnetic moment of this system is obtained by
summing over all the spins in the lattice and is given by 
$\textbf{m}=\sum_{i=1}^{N} \textbf{s}_i$, where $N$ is the number of sites. 
After equilibration, the total magnetic moment of the lattice is recorded at every Monte Carlo step.  The system is cooled from its initial random spin configuration at $T=10$ to $T=0.5$.

\subsection{Site Selection}
\label{s_site-selection}
For every sweep through the lattice of $N$ spins, $N$ spins are offered the chance for reorientation, but only $M$ spins are at different sites.  There are two typical methods of selecting sites.

The ``every-site method" involves giving each site in the lattice one opportunity to reorient, i.e., $M=N$.  This is the method  used in this paper.  In the ``random-site method," $M$ sites are chosen for reorientation at random.  With this method, it is possible to select several sites more than once and some not at all.

We would expect simulations of magnetic noise to produce white noise power spectra at high temperatures.  For the ``every-site method," this is true.  The power spectrum of Heisenberg spins in the high-temperature limit using the ``random-site method" goes as $1/f$, and the $1/f$ noise power spectra is entirely due to the site selection method.  For a given model, the power spectra at different temperatures with the ``random-site method" pivot at a lower frequency compared to the ``every-site method."

\subsection{Time Steps}
\label{s_time-steps}
\begin{figure}
\centering
\includegraphics[width=\linewidth]{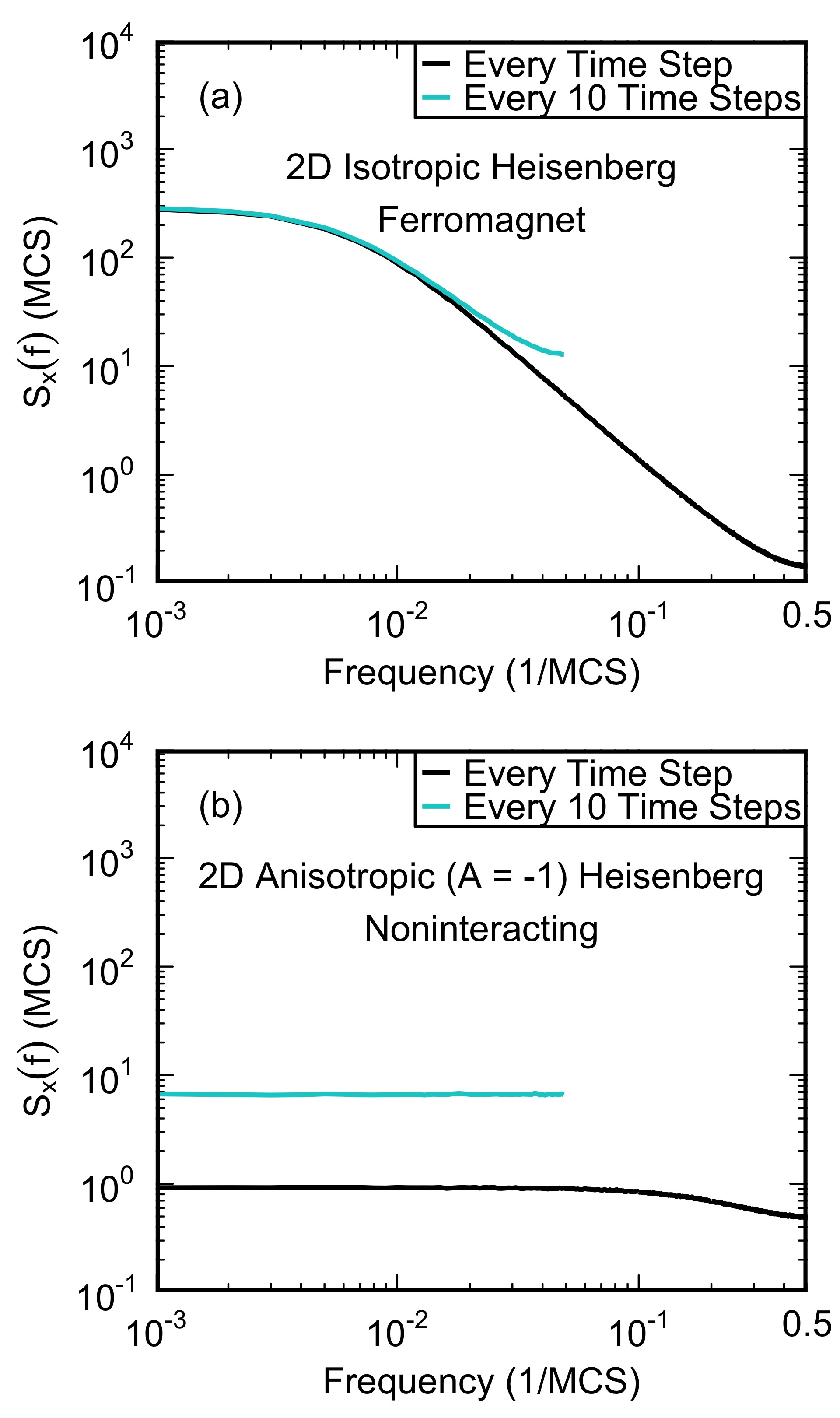}
\caption{Noise power $S_x(f)$ of the x-component of the total magnetic moment versus frequency.  Noise power spectra averaged over 10 segments for the 2D (a) isotropic ($A=0$) Heisenberg ferromagnet ($J_{ij}=1$)  and (b) noninteracting ($J_{ij}=0$), anisotropic ($A=-1$) Heisenberg models from recording the total magnetic moment at every time step and every ten time steps at $T=1$.  The units of noise power are MCS, because the total noise power is dimensionless.}
\label{f_time-steps}
\end{figure}

One time step in a standard Monte Carlo simulation is one Monte Carlo step (MCS).  To simulate a slower recording rate, the total magnetic moment time series can be recorded after several MCS.  A comparison of recording every time step and every 10 time steps is shown in Fig.~\ref{f_time-steps}.  Although the maximum frequency of the power spectra when recording every 10 time steps is smaller than for recording every time step, the total noise power for both cases is equal.  At a given temperature, the total noise power is equal to the variance of the total magnetic moment time series which is dimensionless.  This means that the noise power $S_x(f)$ has units of MCS. Recording every 10 time steps causes the aliasing to occur at a lower frequency, but the location of the low-frequency knee remains unchanged.  The effect is that the pivoting frequency is decreased.  Unless otherwise noted, the time steps are recorded at 1 MCS intervals.

(One might be tempted to record the time series more often than one MCS in an effort to extend to frequencies higher than $0.5 \text{ MCS}^{-1}$. However, in real physical systems, all spins evolve simulataneously so it would be unphysical to record the total magnetic moment before all spins are given the opportunity to reorient.)

\subsection{Equilibration}
\label{s_equilibration}
The test for equilibration follows Bhatt and Young's procedure for the equilibration of Ising spin glasses~\cite{Bhatt1988}.  Two independent replicas of each system with the same exchange couplings and orientation of anisotropic axes are created and run in parallel.  The initial spin configurations for the two replicas are different and random.  For the set of spins \{$\textbf{s}_i$\} with $N$ lattice sites, the spin autocorrelation function for the replica $n$, after an equilibration time $t_0$, is
\begin{equation}
\label{e-Q}
Q^{(n)}(t)=\frac{1}{N} \sum_{i=1}^N \textbf{s}_i^{(n)}(t_0) \cdot \textbf{s}_i^{(n)}(t_0+t),
\end{equation}
where the summation is over all lattice sites.  The spin glass susceptibility for replica $n$ is calculated as the second moment of this overlap and then averaged over 200 different realizations of bonds and anisotropy axes.  This disorder average is denoted by $\left[ \ldots \right]_{\text{av}}$:
\begin{equation}
\label{e-chisg1}
\chi_{\text{SG}}^{(n)}(t)=\frac{1}{N} \left[ \left( \sum_{i=1}^N \textbf{s}_i^{(n)}(t_0) \cdot \textbf{s}_i^{(n)}(t_0+t) \right)^2 \right]_{\text{av}}.
\end{equation}

The equilibration time $t_0$ is chosen from the sequence $1,3,10,30,100,300, \ldots{ }$, etc.  The idea is to compare $\textbf{s}_i^{(n)}(t_0)$ to $\textbf{s}_i^{(n)}(t_0+t)$ as ${t \to \infty}$ to see whether $\textbf{s}_i^{(n)}(t_0+t)$ has lost its ``memory" of $\textbf{s}_i^{(n)}(t_0)$.  In practice, the comparison is done as ${t \to 3 t_0}$.  The spin glass susceptibility in Eq.~(\ref{e-chisg1}) is averaged over a length of time $t_0$:  
\begin{equation}
\label{e-chisg3}
\chi_{\text{SG}}^{(n)}=\frac{1}{N t_0} \left[ \sum_{t=t_0}^{2t_0-1} \left( \sum_{i=1}^N \textbf{s}_i^{(n)}(t_0) \cdot \textbf{s}_i^{(n)}(t_0+t) \right)^2 \right]_{\text{av}}.
\end{equation}

\noindent The summation over $t$ starts at $t_0$ so that the distribution of $Q^{n}(t)$ is Gaussian.  The correlation of the spins at shorter times makes the distribution deviate from a Gaussian.

For small values of $t_0$ and when the system is at low temperatures, there are few spin fluctuations, so $Q^{(n)}(t) \sim 1$ and $\chi_{\text{SG}}^{(n)}(t) \sim N$.  This is in agreement with simulations.

We can also calculate $\chi^{(n)}_{\text{SG}}$ in the high-temperature limit.  We start with two Heisenberg spins $\textbf{s}_1$ and $\textbf{s}_2$ of length 1 that represent $\textbf{s}_i^{(n)}(t_0)$ and $\textbf{s}_i^{(n)}(t_0+t)$, respectively, in Eq.~(\ref{e-chisg3}).  Without loss of generality, we choose one spin to be along the z-axis.  The average square of the dot product is calculated as
\begin{equation}
\label{e-ssah1}
\begin{split}
\langle \left( \textbf{s}_1 \cdot \textbf{s}_2 \right)^2 \rangle_{\text{Heis.}} &= \langle s_1^2 s_2^2 \cos^2(\theta) \rangle \\
&= \langle \cos^2(\theta) \rangle,
\end{split}
\end{equation}
where $\theta$ is the angle between $\textbf{s}_1$ and $\textbf{s}_2$.  The angle $\theta$ is also the polar angle of the spin.  In our simulations, $\cos(\theta)$ is chosen from a uniform distribution, so Eq.~({\ref{e-ssah1}) can be simplified:
\begin{equation}
\label{e-ssah2}
\begin{split}
\langle \left( \textbf{s}_1 \cdot \textbf{s}_2 \right)^2 \rangle_{\text{Heis.}} &= \langle \cos^2 (\theta) \rangle \\
&= \frac{1}{2} \int_{-1}^1 \cos^2(\theta) d(\cos(\theta)) \\
&= \frac{1}{3}
\end{split}
\end{equation}

We repeat the process for an XY spin with an angle $\phi$ between spins that is chosen from a uniform distribution:
\begin{equation}
\label{e-ssax}
\begin{split}
\langle \left( \textbf{s}_1 \cdot \textbf{s}_2 \right)^2 \rangle_{\text{XY}} &= \langle s_1^2 s_2^2 \cos^2 (\phi) \rangle \\
&= \langle \cos^2 (\phi) \rangle \\
&= \frac{1}{2 \pi} \int_0^{2 \pi} \cos^2(\phi) d \phi \\
&= \frac{1}{2}.
\end{split}
\end{equation}

For Ising spins,
\begin{equation}
\label{e-ssai}
\begin{split}
\langle \left( \textbf{s}_1 \cdot \textbf{s}_2 \right)^2 \rangle_{\text{Ising}} &= \langle \left( \pm s_1 s_2 \right)^2 \rangle \\
&= \langle s_1^2 s_2^2 \rangle \\
&= 1.
\end{split}
\end{equation}

From Eqs.~(\ref{e-ssah2}), (\ref{e-ssax}), and (\ref{e-ssai}), we can see that for spins with $m$ components,
\begin{equation}
\label{e-ssm}
\langle \left( \textbf{s}_1 \cdot \textbf{s}_2 \right)^2 \rangle = \frac{1}{m}.
\end{equation}

Combining Eq.~(\ref{e-ssm}) with Eq.~(\ref{e-chisg3}), for high temperatures, we get $\chi^{(n)}_{\text{SG}} = \frac{1}{m}$ which is seen in simulations.



We then define the average of the two single replica susceptibilities
\begin{equation}
\label{e-chisga}
\overline{\chi}_{\text{SG}}=\frac{1}{2}\left( \chi_{\text{SG}}^{(1)}+ \chi_{\text{SG}}^{(2)} \right)
\end{equation}
\noindent as the two times spin glass susceptibility.

The spin glass susceptibility may also be calculated from the spin overlap of the two different replicas.  The mutual overlap between the spins $\textbf{s}_i^{(1)}$ and $\textbf{s}_i^{(2)}$ of the two replicas is 
\begin{equation}
\label{e-Q'}
Q(t)=\frac{1}{N} \sum_{i=1}^N \textbf{s}_i^{(1)}(t_0+t) \cdot \textbf{s}_i^{(2)}(t_0+t).
\end{equation}

The spin glass susceptibility is calculated from the spin overlap:
\begin{equation}
\label{e-chisg4}
\chi_{\text{SG}}=\frac{1}{N t_0} \left[ \sum_{t=t_0}^{2t_0-1} \left( \sum_{i=1}^N \textbf{s}_i^{(1)}(t_0+t) \cdot \textbf{s}_i^{(2)}(t_0+t) \right)^2 \right]_{\text{av}}.
\end{equation}

For all temperatures, as the equilibration time is approached, the spin glass susceptibilities converge; the two times susceptibility $\overline{\chi}_{\text{SG}}$ (Eq.~\ref{e-chisga}) approaches the true susceptibility from above and the replica susceptibility $\chi_{\text{SG}}$ (Eq.~\ref{e-chisg4}) from below.

After sufficiently long equilibration times, $\chi_{\text{SG}}$ and $\overline{\chi}_{\text{SG}}$ agree.  We define the system to be equilibrated if
\begin{equation}
\Delta \chi_{SG} =  \lfrac{|\chi_{\text{SG}}-\overline{\chi}_{\text{SG}}|}{\frac{1}{2} \left( \chi_{\text{SG}}+\overline{\chi}_{\text{SG}} \right) } 
\end{equation}
is less than 5\% for three consecutive times in the $t_0$ sequence; then we declare it equilibrated at the fourth time.  For example, if the last three equilibration times are $t_1=3000$, $t_2=10000$, $t_3=30000$, then the equilibration time $t_4=100000$.  At each temperature, the initial equilibration time is $1 \times 10^5$ MCS, and it is increased if the system is not equilibrated.

\subsection{Noise Power}

The time series of each component of the total magnetic moment is given by $m_a(t_j)$, where $a=x,y,z$.  The deviation from the average is $\delta m_a(t_j) = m_a(t_j) - \overline{m_a(t_j)}$.  The noise power spectral density can be determined from the Fourier transform of the autocorrelation function of the time series.  The autocorrelation is given by
\begin{equation}
\label{e-C1}
C_a(t_k)=\frac{1}{N_\tau} \sum_{j=0}^{N_{\tau}-1} \delta m_a(t_j) \delta m_a^*(t_{j}-t_{k}),
\end{equation}
where $\delta m_a^*(t_j-t_k)$ is the complex conjugate of $\delta m_a(t_j-t_k)$.  For a discrete time series $m_a(t_j)$ of length $N_\tau$, the discrete inverse Fourier transform is given by
\begin{equation}
\label{e-dft}
\delta m_a(t_j)=\frac{1}{N_\tau} \sum_{k=0}^{N_\tau-1} \tilde{m}_a(f_k) e^{2\pi i f_k t_j /N_\tau},
\end{equation}
where $\tilde{m}_a(f_k)$ is the Fourier transform of the time series.  Using Eq.~(\ref{e-dft}), the autocorrelation function becomes
\begin{multline}
C_a(t_k)=\frac{1}{N_\tau^3} \sum_{j=0}^{N_{\tau}-1} \left[ \left( \sum_{l=0}^{N_\tau-1} \tilde{m}_a(f_l) e^{2\pi i f_l t_j/N_\tau} \right) \right. \times \\
\left. \left( \sum_{n=0}^{N_\tau-1} \tilde{m}^*_a(f_n) e^{-2\pi i f_n (t_j-t_k)/N_\tau} \right) \right].
\end{multline}

This can be simplified to
\begin{equation}
\label{e-autocorrelation}
C_a(t_k)=\frac{1}{N_\tau^2} \sum_{n=0}^{N_{\tau}-1} | \tilde{m}_a(f_n)|^2 e^{2\pi i f_n t_k/N_\tau}.
\end{equation}
Taking the Fourier transform of Eq.~(\ref{e-autocorrelation}) gives the periodogram estimate~\cite{Press2007} for computing the noise power $S_a(f_k)$ for the axes $a=x,y,z$:
\begin{equation}
\label{e-noise_power}
S_a(f_k)=\frac{1}{N_\tau} \frac{1}{N} |\tilde{m}_a(f_k)|^2,
\end{equation}
where the power spectra is divided by $N$ so that $S_a(f_k)$ is the noise power per site for spin component $a$.  The power spectra are normalized so that at a given temperature, the total noise power is equal to the variance of the total magnetic moment time series divided by the number of sites.

The Fourier transform $\tilde{m}_a(f_k)$ is computed using the C subroutine library FFTW~\cite{Frigo2005}.  At a given temperature, the time series $\delta m_a(t_j)$ for a given run is split into either 10 or 100 segments (blocks) of equal length.  The power spectrum is found for each segment and is averaged over these segments to give a smoother power spectrum.  At each temperature, the spectra are averaged over 200 independent runs.  The power spectra used in all plots shown in this paper are averaged over 100 segments.

\section{Results}
\label{s_results}

\subsection{Low-Frequency Knee}

At lower frequencies of the power spectra, there are ``knees" where the power spectra transition from $A^2/f^\alpha$ to white noise at low frequency as shown in Fig.~\ref{f_knee}.  The frequency of the knee decreases as the temperature decreases.

\begin{figure}
\centering
\includegraphics[width=\linewidth]{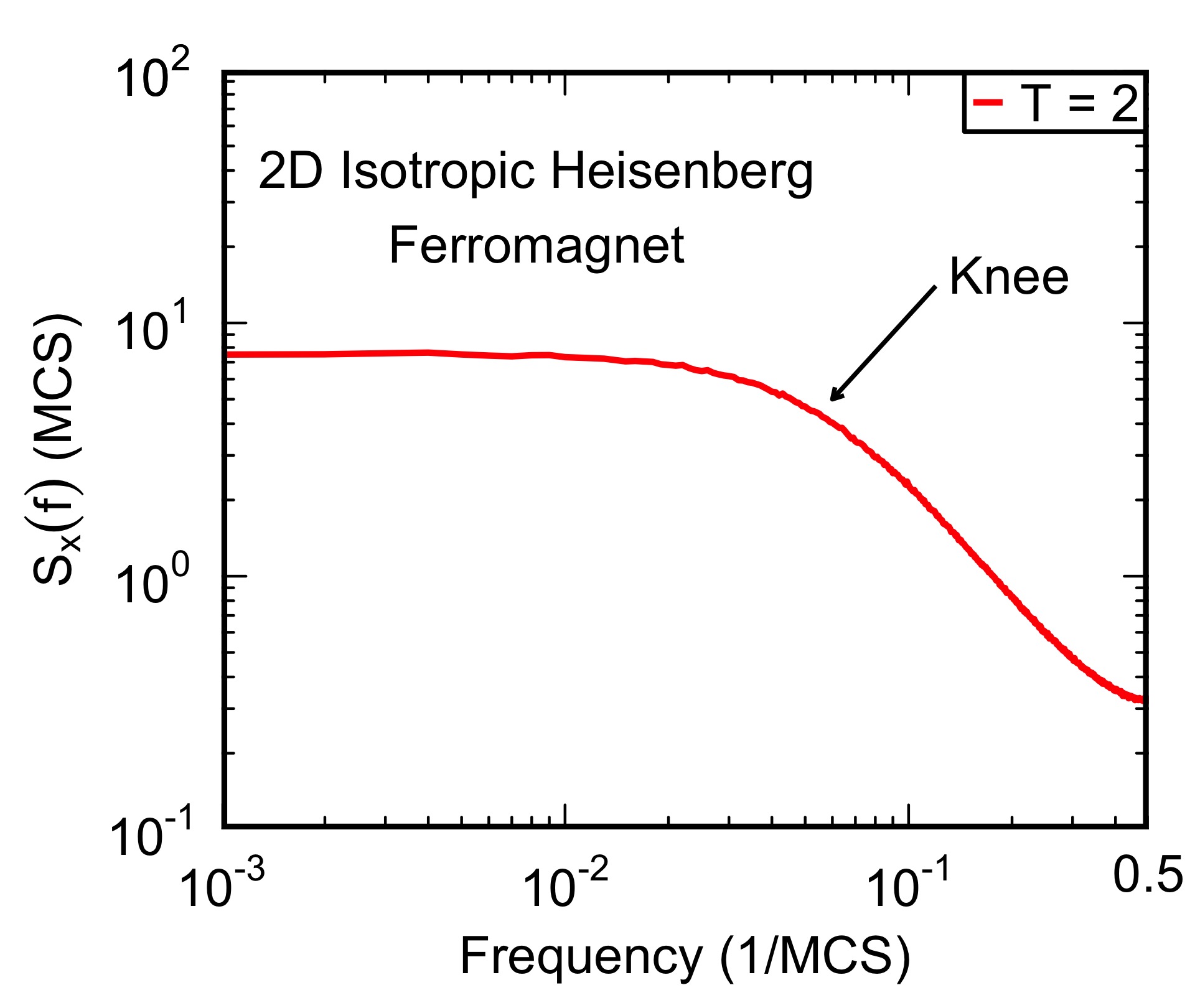}
\caption{Noise power $S_x(f)$ of the x-component of the total magnetic moment of a 2D isotropic ($A=0$) Heisenberg ferromagnet ($J_{ij}$=1) versus frequency.  The low-frequency knee of the power spectrum averaged over 100 segments is shown for $T=2$.}
\label{f_knee}
\end{figure}

The low-frequency knee is due to finite size effects of the lattice~\cite{Chen2007}.  The correlation length increases as the temperature approaches the transition temperature $T_c$ associated with the relevant order parameter.  The relaxation time increases as the correlation length increases, so the system takes longer to equilibrate.  In previous simulations of the ferromagnetic Ising model and 5-state Potts model, it was found that the knee frequency is proportional to the system's minimum relaxation rate $\tau^{-1}$, i.e., the system's maximum relaxation time ~\cite{Chen2007}.  This means that as the temperature decreases and approaches $T_c$, the knee frequency decreases.  Since the relaxation time increases with system size, we expect the knee frequency to decrease with system size~\cite{Chen2007}.  (So in real materials or devices where the lattice size is huge, one does not see a knee in the noise power spectrum.) One should look at the noise in the relevant order parameter to see the knee frequency decrease.  The order parameters are magnetization for the ferromagnetic systems, staggered magnetization for the antiferromagnetic systems, and the spin glass order parameter $q$ for spin glass systems.  We investigate this below.

According to the Mermin-Wagner-Hohenberg theorem~\cite{Mermin1966,Hohenberg1967}, two-dimensional Heisenberg spin systems should not undergo a phase transition.  The order parameters of the ferromagnet, antiferromagnet, and spin glass systems show that Heisenberg systems exhibit ordering at a positive $T_c$.  This discrepancy is because the Mermin-Wagner-Hohenberg theorem holds in the thermodynamic limit, while these simulations are for finite lattices.  Simulations show that as the system size increases, the value of $T_c$ decreases toward zero.  This system size dependence is shown in Fig.~\ref{f_size-dependence}.  We define the magnetic susceptibility of the total magnetic moment per site as
\begin{equation}
\chi_{\rm total} = \frac{1}{k_B T}\left( \left\langle \left| \frac{m}{N} \right|^2 \right\rangle - \left\langle \left|\frac{m}{N}\right| \right\rangle^2 \right),
\end{equation}
where we set $k_B=1$, $T$ is temperature, $N$ is the number of lattice sites, and the magnitude of the total magnetization $\left| m \right| = \sqrt{m_x^2+m_y^2+m_z^2}$. As the system size increases, the peaks in the magnetic susceptibility occur at lower temperatures.

\begin{figure}
\includegraphics[width=\linewidth]{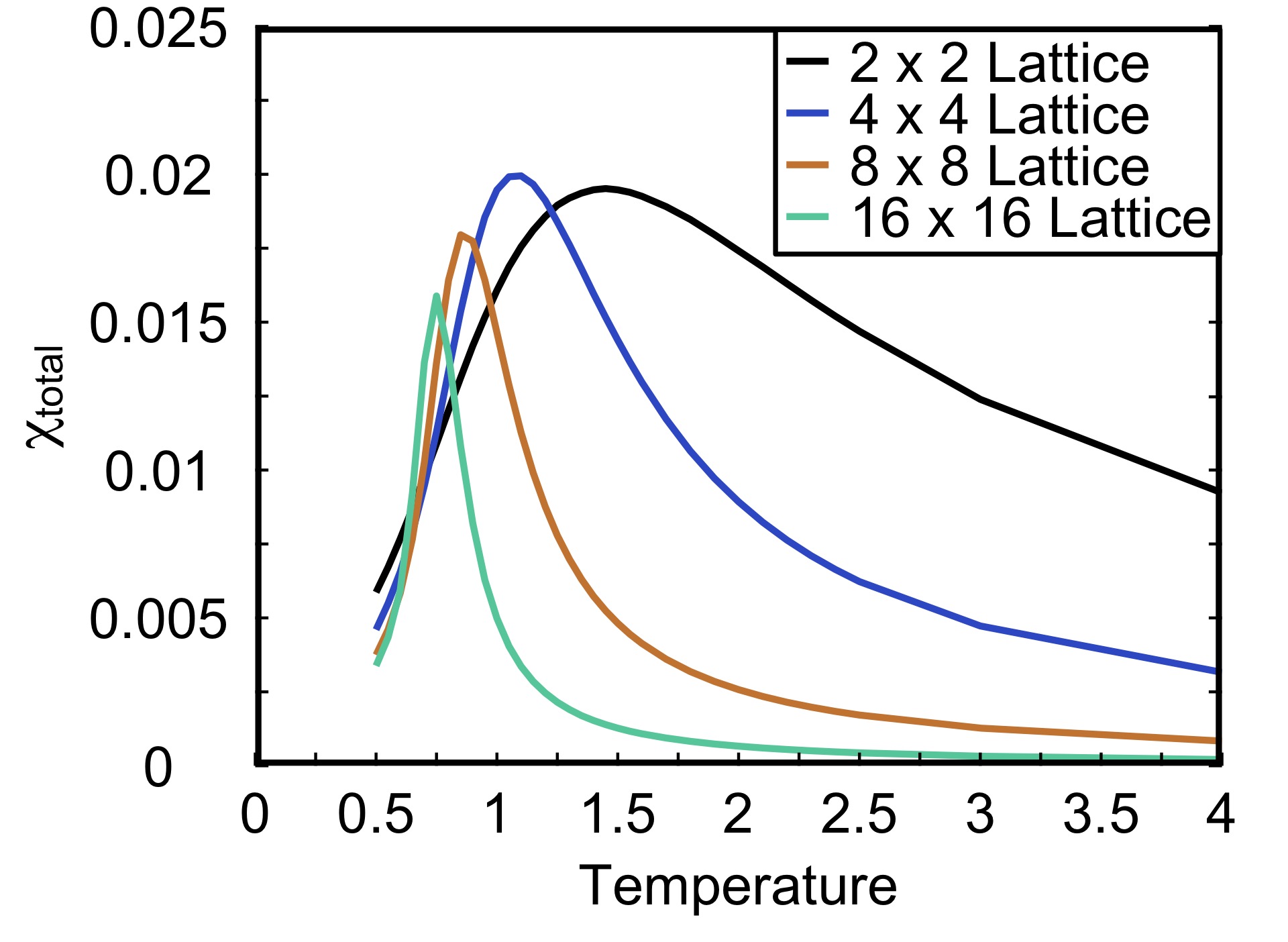}
\caption{Magnetic susceptibility of the magnitude of the total magnetic moment per site versus temperature for the 2D Heisenberg Ferromagnet as a function of system size.}
\label{f_size-dependence}
\end{figure}

\subsection{Aliasing}
\label{aliasing}

At higher frequencies near $0.5 \text{ MCS}^{-1}$, there is an upturn of the noise power due to aliasing that is shown in Fig.~\ref{f_regions}.  The aliasing is due to the periodicity of the factor $e^{-2\pi i f t}$ used in the Fourier transform~\cite{Press2007}.  When calculating the discrete Fourier transform, frequency components of the power spectra that are greater than $0.5 \text{ MCS}^{-1}$ are translated into the range $0 \text{ MCS}^{-1} < f < 0.5 \text{ MCS}^{-1}$~\cite{Press2007}.  As we explain later in this section, under certain conditions, this aliasing can cause spectral pivoting where the power spectra at different temperatures cross within a narrow range of frequency.

\begin{figure}
\centering
\includegraphics[width=\linewidth]{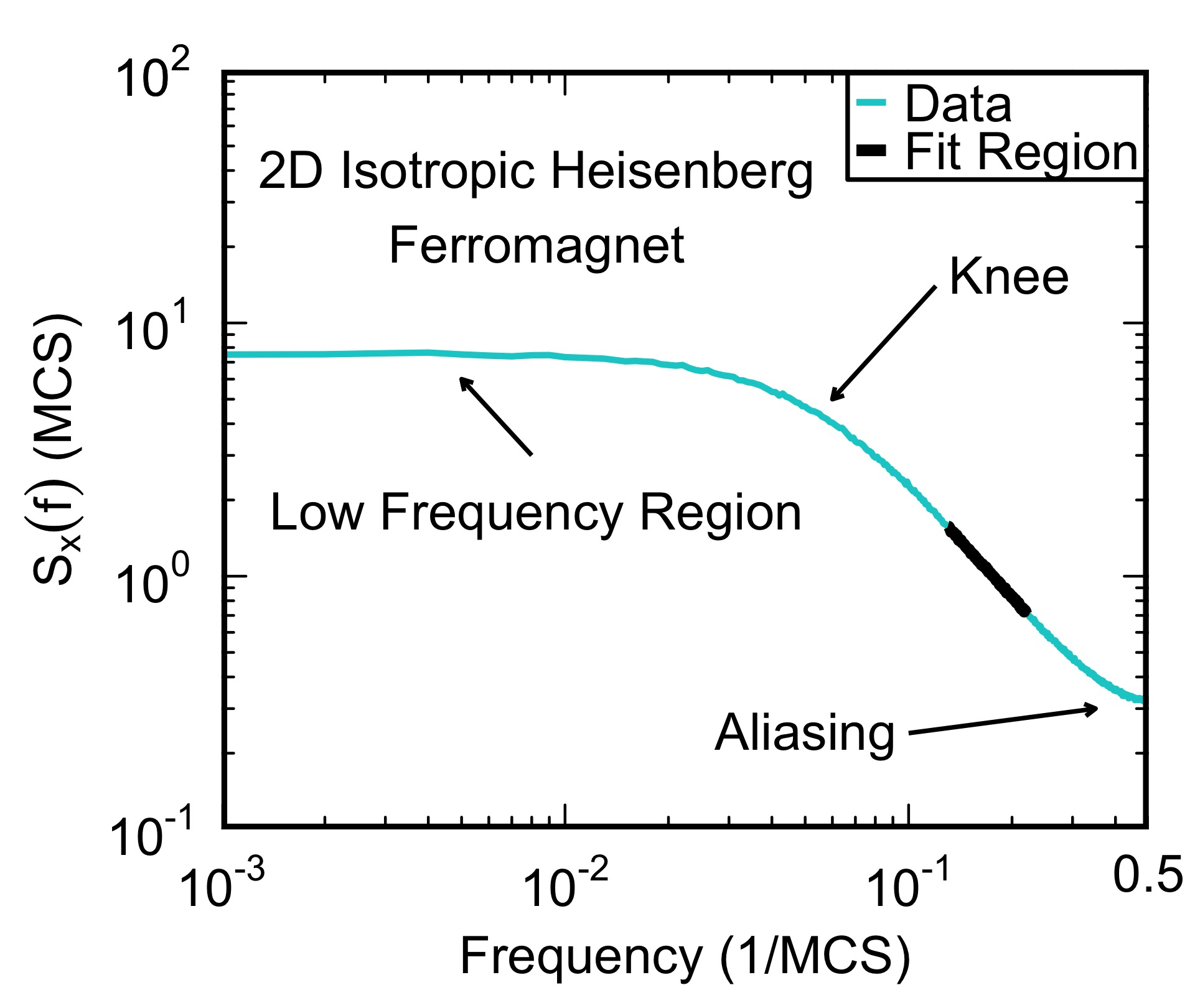}
\caption{Noise power $S_x(f)$ of the x-component of the total magnetic moment of a 2D isotropic ($A=0$) Heisenberg ferromagnet ($J_{ij}$=1) versus frequency.  Labeled regions of the power spectrum averaged over 100 segments at $T=2$.}
\label{f_regions}
\end{figure}


\subsection{Noise Exponents}
\label{exponents}

To determine the noise amplitude ($A^2$) and the noise exponent ($\alpha$), the function $A^2/f^\alpha$ is fit to the region of the power spectra that is linear on a log-log plot and that lies between the low-frequency knee and high-frequency upturn due to aliasing as shown in Fig.~\ref{f_regions}.  More details about this fitting process can be found in Appendix~\ref{a_fitting}.  In performing fits, $A^2/f^\alpha$ is fit to the 10-averaged power spectra, so the fits are linear on log-log plots.  Each x, y, and z component of the spin results in a power spectrum, and the amplitudes and exponents are determined for each component.  The noise amplitudes and exponents as a function of temperature are shown in Figs.~\ref{f_noise-amplitudes} and \ref{f_noise-exponents}, respectively.  In Fig.~\ref{f_noise-exponents}, the shaded region indicates the experimental range of noise exponents where $0.5 \leq \alpha \leq 1$ and $1 \leq T \leq 2$.  (The lower bound of $T \geq 1$ is set by the energy scale of the exchange constant $J_{ij}$.)  From the plot of noise exponents, we can see that the spin glass systems are the most consistent with experiment.  

\begin{figure}
\includegraphics[width=\linewidth]{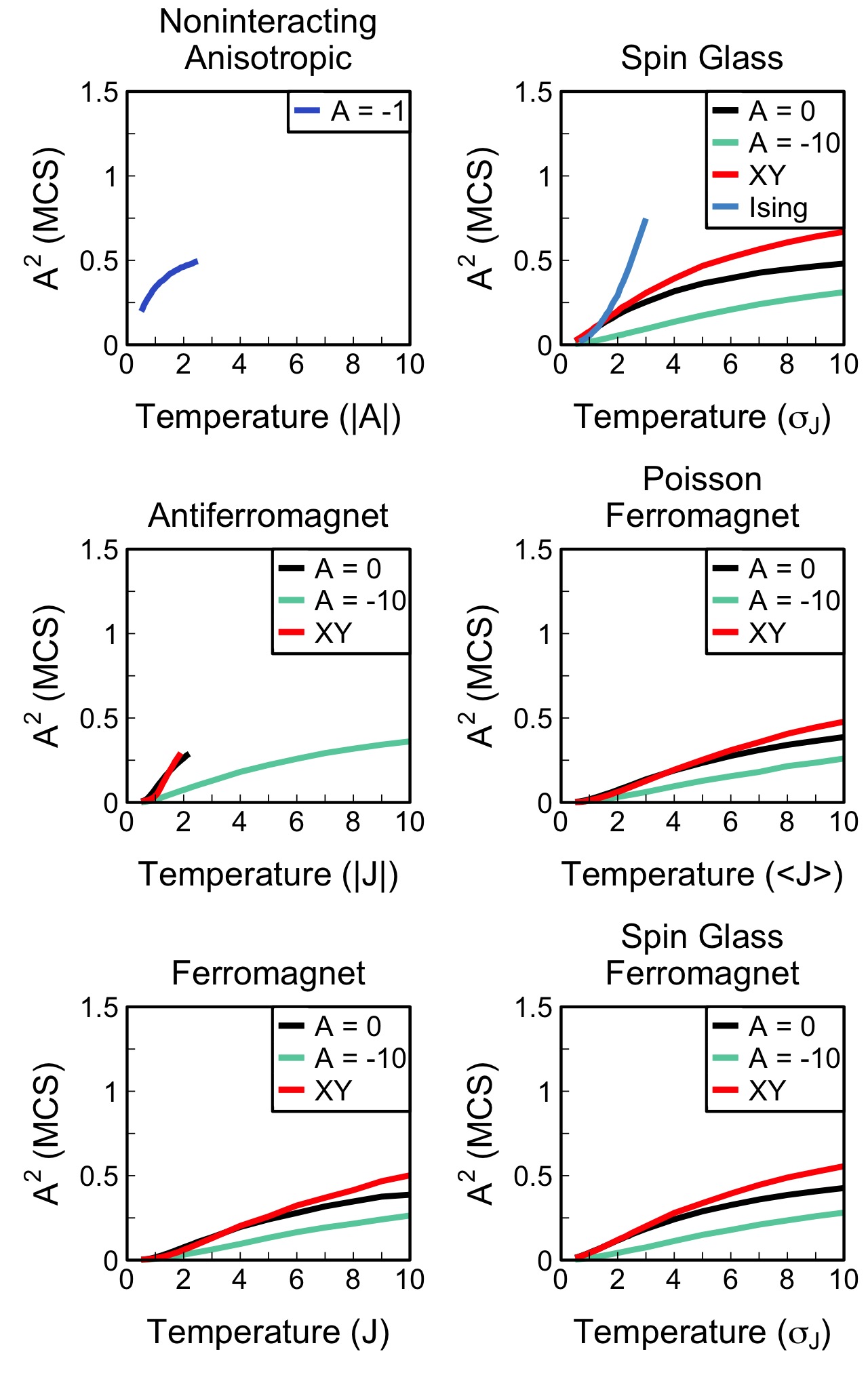}
\caption{Noise amplitude averaged over spin components as a function of temperature for ($0.5 \leq T \leq 10$).}
\label{f_noise-amplitudes}
\end{figure}

\begin{figure}
\includegraphics[width=\linewidth]{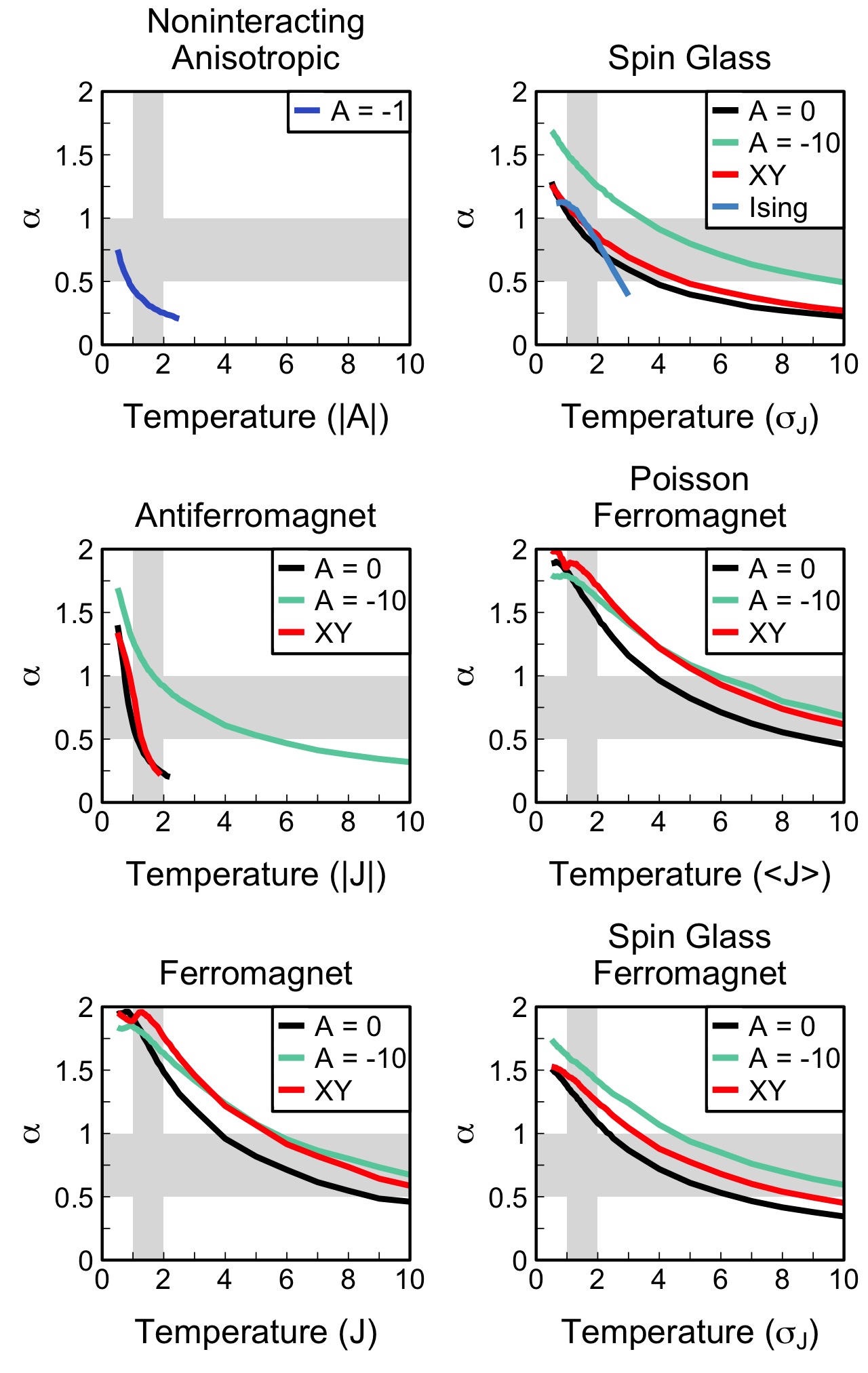}
\caption{Noise exponents averaged over spin components as a function of temperature for ($0.5 \leq T \leq 10$).  The shaded region indicates the experimental range of $0.5 \leq \alpha \leq 1$ and $1 \leq T \leq 2.$}
\label{f_noise-exponents}
\end{figure}

\subsection{Pivoting}
\label{s_pivoting}

If the knee and aliasing upturn are close in frequency, then the two regions around these features overlap, and the frequency range of the power law fit is reduced.  This effect is shown in Fig.~\ref{f_fit-range} for the isotropic Heisenberg ferromagnet at $T=10$.

\begin{figure}
\centering
\includegraphics[width=\linewidth]{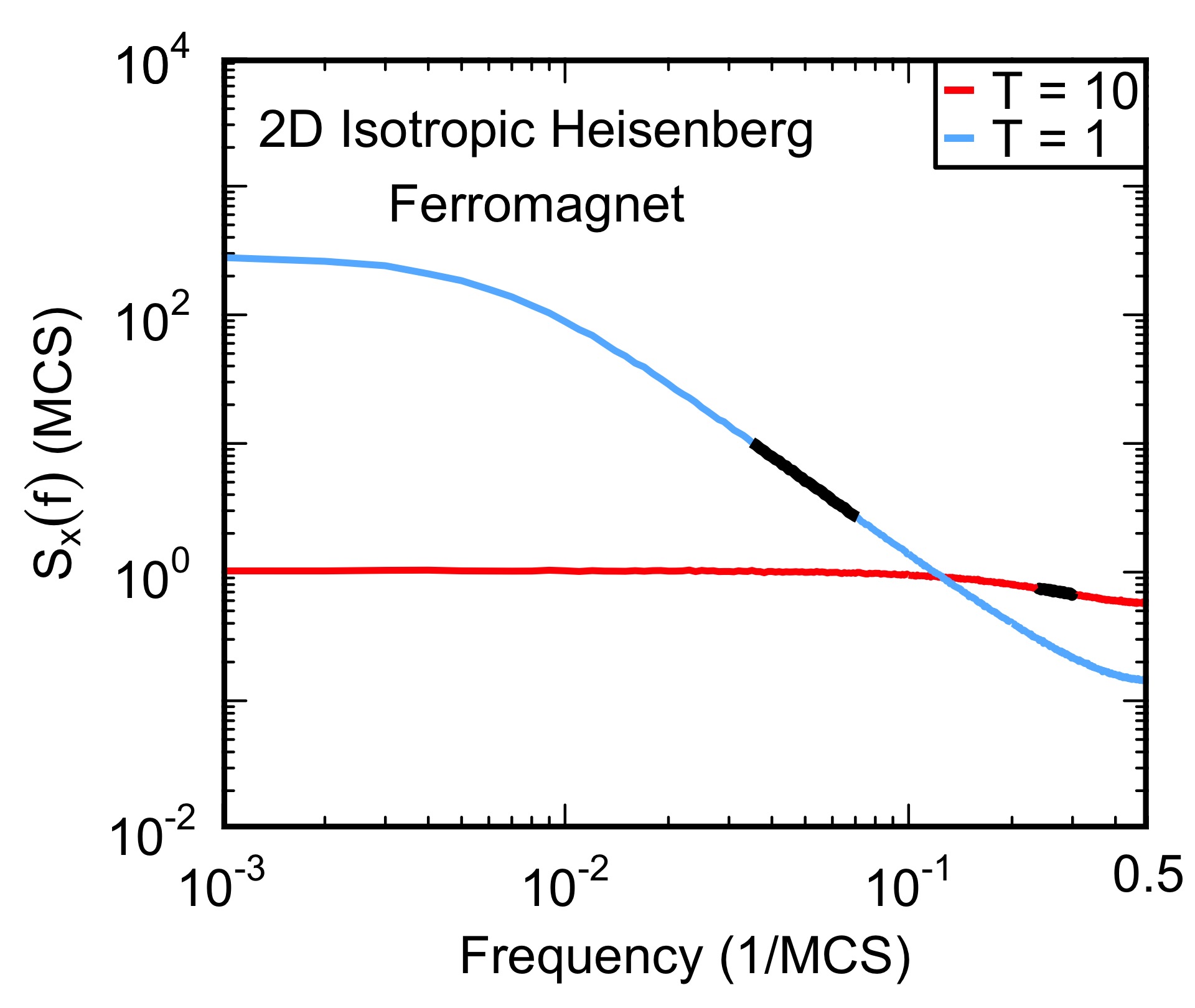}
\caption{Noise power $S_x(f)$ of the x-component of the total magnetic moment of a 2D isotropic ($A=0$) Heisenberg ferromagnet ($J_{ij}=1$) versus frequency.  Black lines indicate the frequency fit range of the noise power spectra averaged over 100 segments at $T=10$ and $T=1$.}
\label{f_fit-range}
\end{figure}

Another effect of the aliasing upturn being close to the knee is that the power spectra at different temperature pivot about a common frequency as seen in Figs.~\ref{f_pivot-all} and \ref{f_pivot-ferromagnet}.  In Fig.~\ref{f_pivot-all}, the noise power as a function of frequency is shown for different temperatures for the isotropic ferromagnetic, isotropic antiferromagnetic, isotropic spin glass, and noninteracting anisotropic $(A=-1)$ Heisenberg spin models.  Since the power spectra for the antiferromagnetic spin system do not cross at a common frequency, they do not pivot.  The power spectra for the isotropic ferromagnetic, isotropic spin glass, and noninteracting, anisotropic $(A=-1)$    systems pivot at high temperature.  To examine this further, in Fig.~\ref{f_pivot-ferromagnet}, the noise power as a function of frequency is shown for the Heisenberg ferromagnet for $0.5 \leq T \leq 10$.  Over this large range of temperature, the power spectra does not pivot.  The inset of Fig.~\ref{f_pivot-ferromagnet} shows pivoting for the high-temperature range $4 \leq T \leq 10$.  Fig.~\ref{f_pivot-ising-low} shows the pivoting of the power spectra for the 2D Ising spin glass for $1.7 \leq T \leq 2.3$.    The anomalous high-temperature pivoting of power spectra for $4 \leq T \leq 10$ is presented in Appendix~\ref{a_ising-pivot} for the 2D Ising spin glass.

\clearpage
\onecolumngrid

\begin{figure}[p]
\centering
\includegraphics[width=\linewidth]{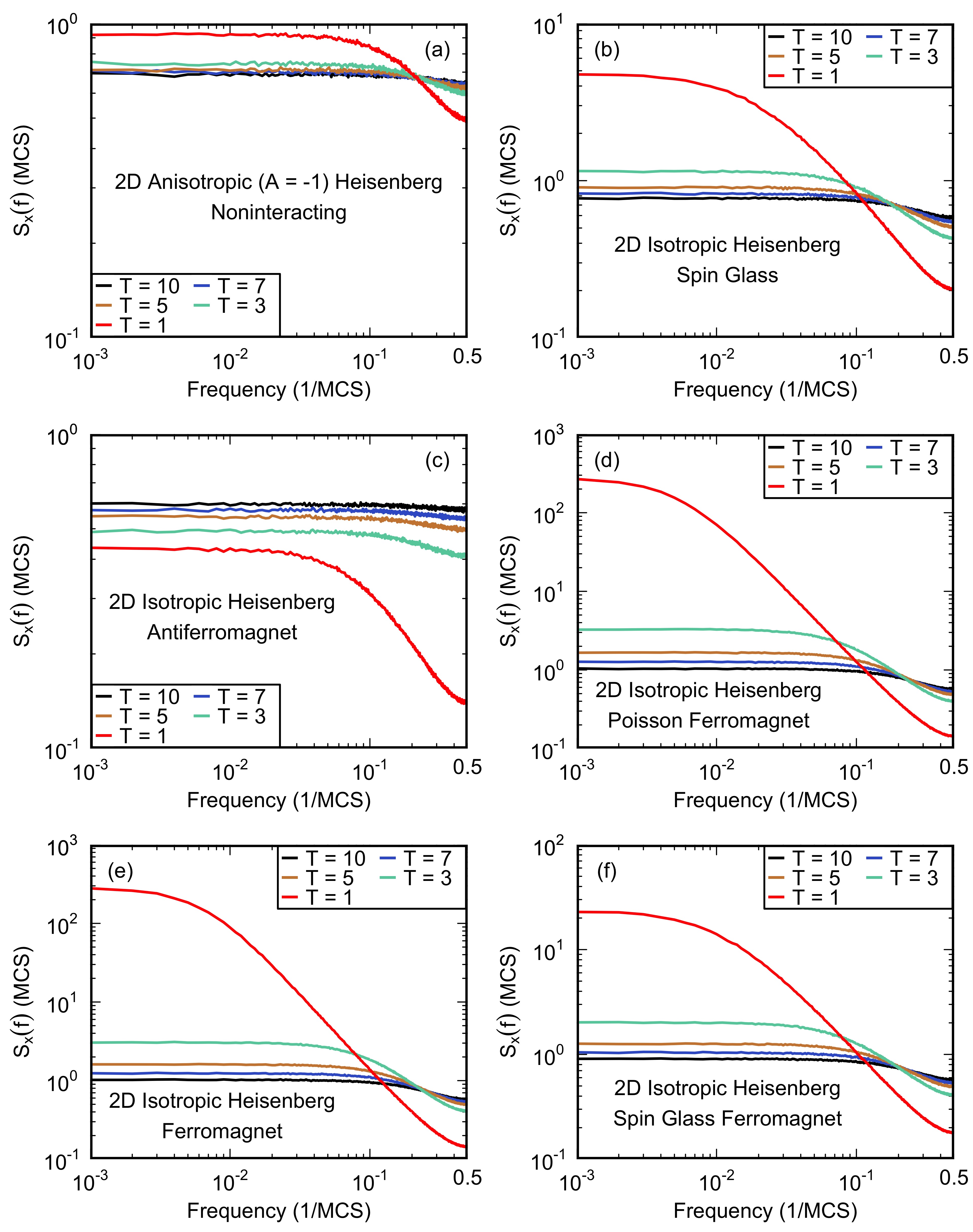}
\caption{Noise power $S_x(f)$ of the x-component of the total magnetic moment versus frequency.  Spectral pivoting of power spectra averaged over 100 segments of the 2D
(a) noninteracting ($J_{ij}=0$) anisotropic ($A=-1$),
(b) isotropic ($A=0$) spin glass ($\langle J_{ij} \rangle=0, \sigma_{J_{ij}}=1$),
(c) isotropic ($A=0$) antiferromagnetic ($J_{ij}=-1$),
(d) isotropic ($A=0$) Poisson ferromagnetic ($\langle J_{ij} \rangle=1$, $\sigma^2_{J_{ij}}=0.2$),
(e) isotropic ($A=0$) ferromagnetic ($J_{ij}=1$), and
(f) isotropic ($A=0$) spin glass ferromagnetic ($\langle J_{ij} \rangle=0.5, \sigma_{J_{ij}}=1$) Heisenberg models for $1 \leq T \leq 10$.}
\label{f_pivot-all}
\end{figure}

\clearpage
\twocolumngrid
\clearpage
\vfill

\begin{figure}
\centering
\includegraphics[width=\linewidth]{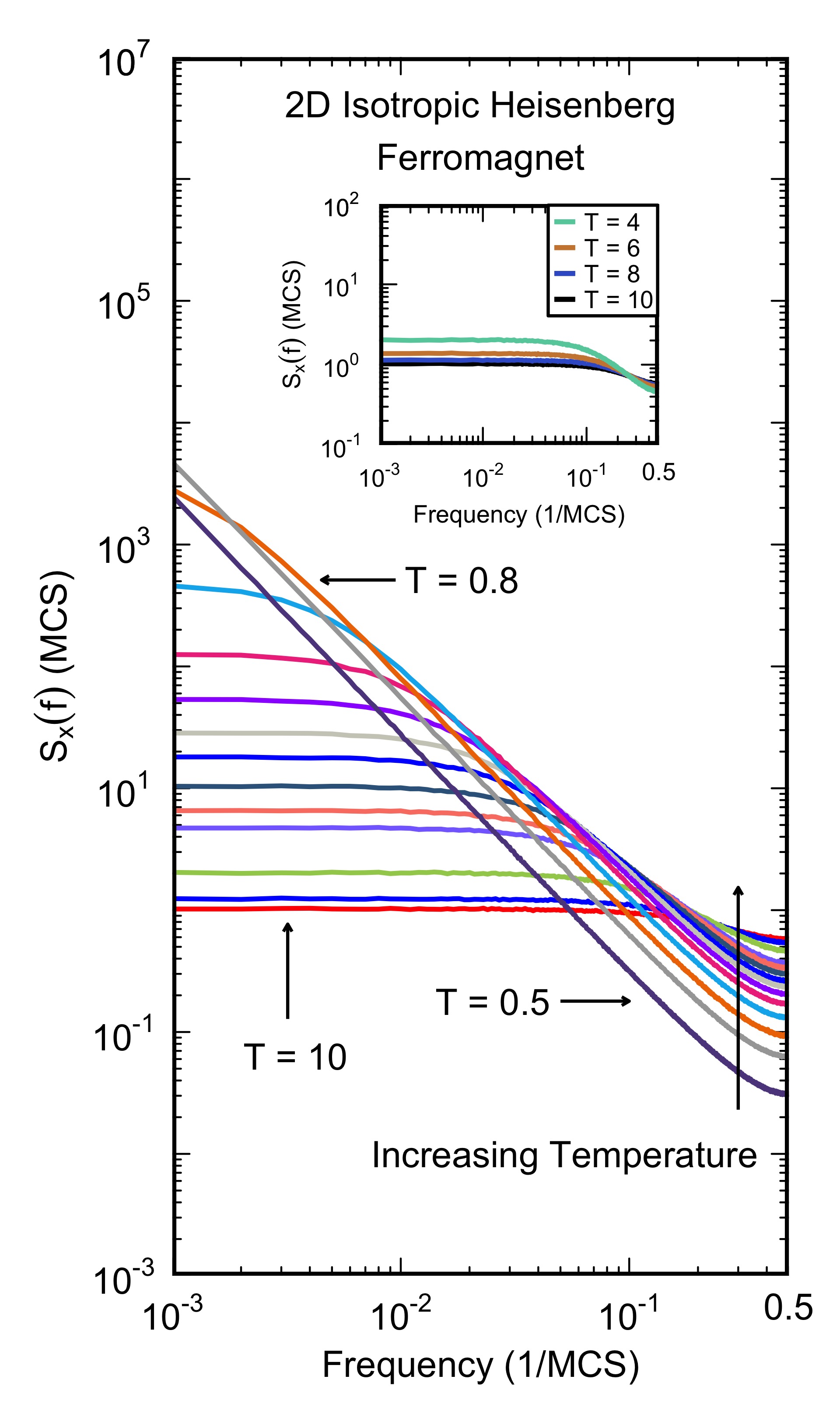}
\caption{Noise power $S_x(f)$ of the x-component of the total magnetic moment of a 2D isotropic $(A=0)$ Heisenberg ferromagnet ($J_{ij}=0$) versus frequency.  Spectral pivoting of the power spectra averaged over 100 segments for $0.5 \leq T \leq 10$.  The inset shows the power spectra averaged over 100 segments for $4 \leq T \leq 10$.}
\label{f_pivot-ferromagnet}
\end{figure}

\begin{figure}
\centering
\includegraphics[width=\linewidth]{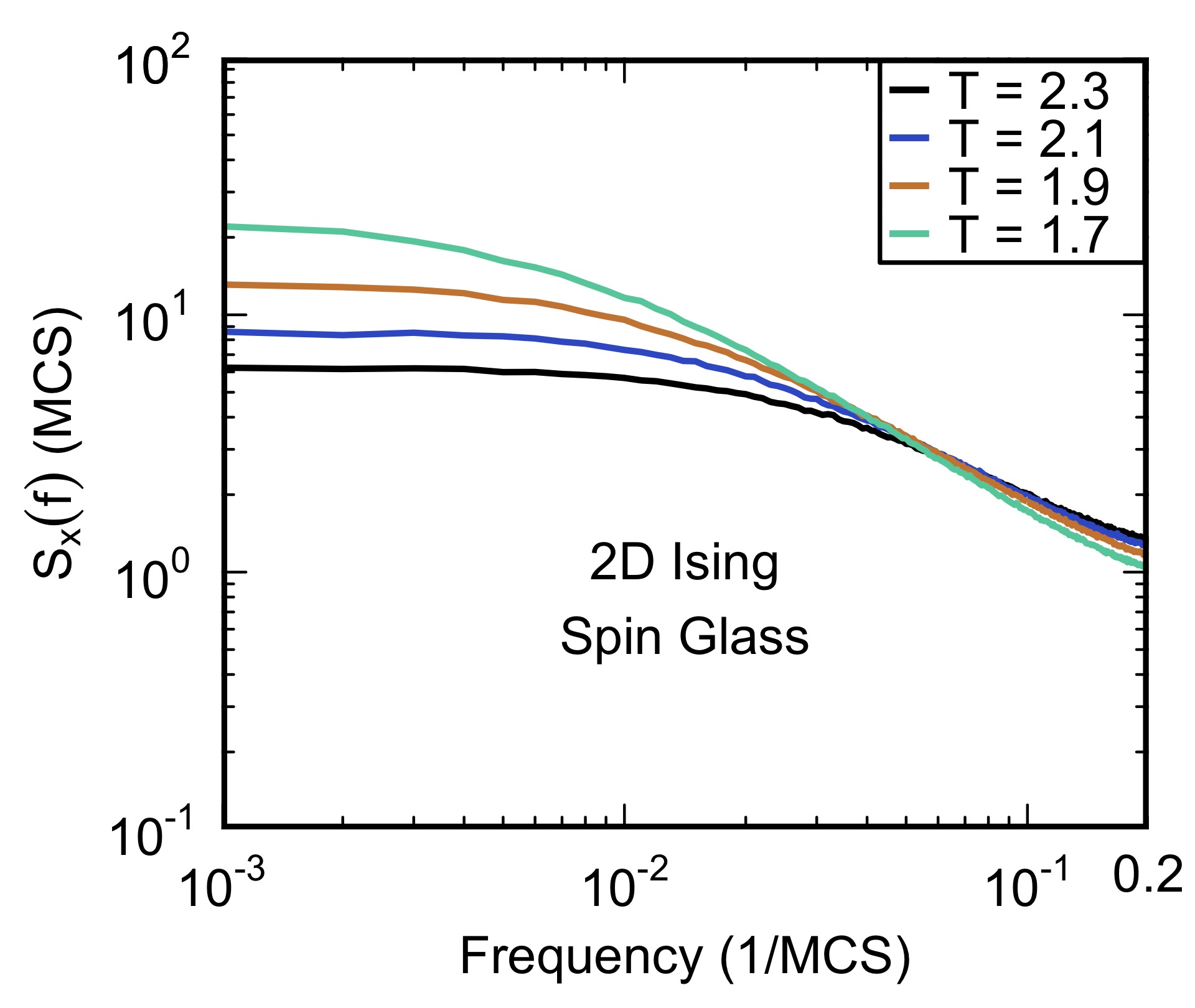}
\caption{Noise power $S(f)$ of the total magnetic moment of a 2D Ising spin glass ($\langle J_{ij} \rangle=0, \sigma_{J_{ij}}=1$) versus frequency.  Spectral pivoting of power spectra averaged over 100 segments for $1.7 \leq T \leq 2.3$.}
\label{f_pivot-ising-low}
\end{figure}

Spectral pivoting occurs for all systems tested except for the Heisenberg antiferromagnets with $A=0$ and $A=-1$ as well as the XY antiferromagnet (see Fig.~\ref{f_pivot-all}).  Spectral pivoting is most evident at high temperatures well above the magnetic transition temperature $T_c$. This is consistent with experiments on flux noise where there is no conclusive evidence for a magnetic phase transition, indicating that the experimentally observed pivoting occurs at temperatures above any magnetic phase transition. 

For our simulated spin systems that exhibit pivoting, as their temperature decreases, the spins change their orientations more slowly; the low-frequency noise power increases, and the noise power at high frequencies decreases.  Note that the total noise power i.e. the integrated area under spectral density curve, is the same for all temperatures for the noninteracting Heisenberg model with anisotropy $A=-1$ and spin glass models with $T_c=0$ is zero.  As a result, increasing noise power at low frequencies means decreasing noise power at high frequencies.

We find that that the crossing frequency of the power spectra has a weak temperature dependence of the form $f_c=B \cdot T + f_0$.  For noninteracting Heisenberg spins with anisotropy $A=-1$, $B=0.023$ and $f_0=-0.16$ for $0.5 \leq T \leq 2.5$.  For comparison, $B=0.11$ and $f_0=-0.12$ for the isotropic Heisenberg ferromagnet for $1.25 \leq T \leq 2.5$.  Spectral pivoting is an artifact that occurs at high temperature where the knee and aliasing regions are close to each other as seen in the inset of Fig.~\ref{f_pivot-ferromagnet}.  At low temperatures where the knee and aliasing are not close, we do not see pivoting which can be seen in Fig.~\ref{f_pivot-ferromagnet}.

By changing a few simulation parameters, the pivoting can be affected.  The pivoting of the power spectra of the Heisenberg ferromagnet when recording the magnetic moment at every time step using the ``every-site method" is shown in Fig.~\ref{f_pivot-method}(a).  Using the method of randomly selecting sites for reorientation outlined in Sec.~\ref{s_site-selection} results in a lower crossing frequency by lowering the frequency where aliasing occurs as shown in Fig.~\ref{f_pivot-method}(b).  Recording the magnetic moment time series every ten time steps as described in Sec.~\ref{s_time-steps} also lowers the crossing frequency compared to recording the magnetic moment time series at every time step as shown in Fig.~\ref{f_pivot-method}(c).

\begin{figure}
\centering
\includegraphics[width=0.95 \linewidth]{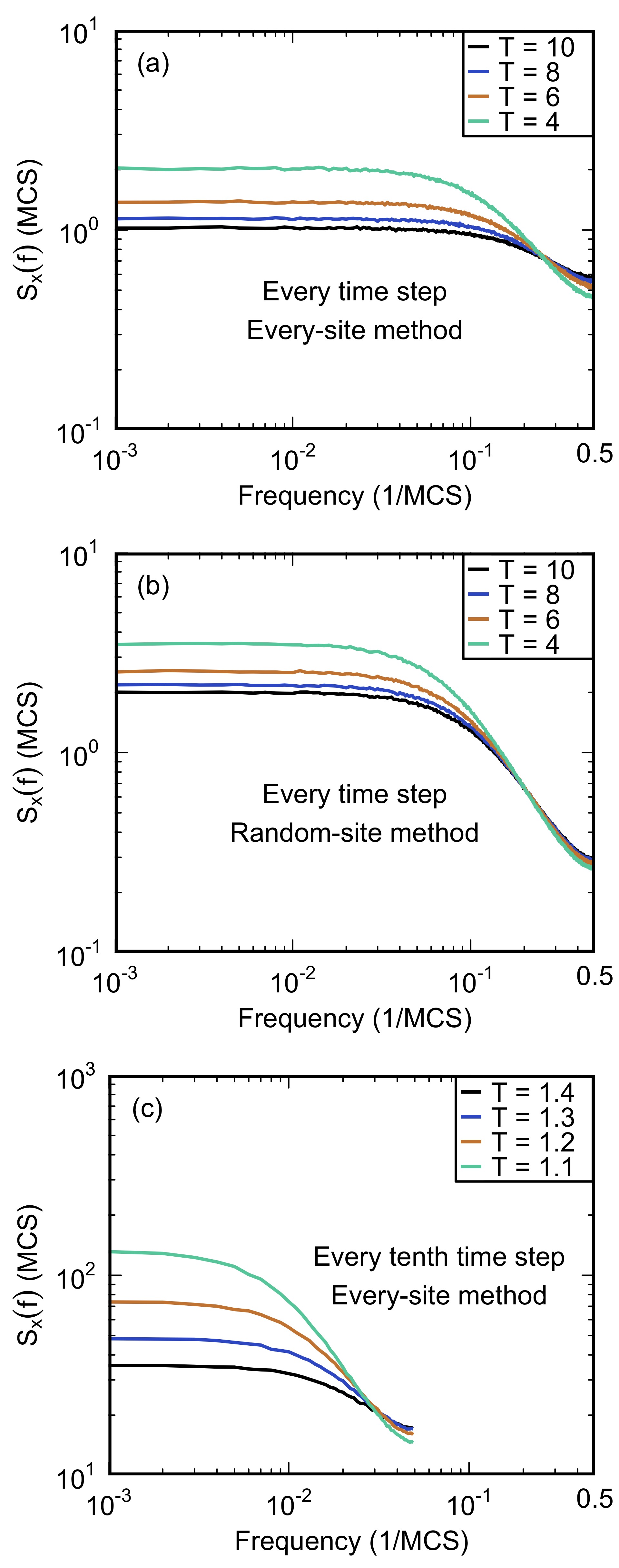}
\caption{Noise power $S_x(f)$ of the x-component of the total magnetic moment of a 2D isotropic ($A=0$) Heisenberg ferromagnet ($J_{ij}=1$) versus frequency.  Spectral pivoting of x-component of the power spectra averaged over 100 segments resulting from recording the magnetic moment time series (a) at every time step using the ``every-site method," (b) at every time step using the ``random-site method," and (c) at every tenth time step using the ``every-site method."}
\label{f_pivot-method}
\end{figure}

\subsection{Mean-Square Flux Noise}
As we mentioned earlier, the mean-square flux noise is given by
\begin{equation}
\langle \Phi^2 \rangle = \int_{f_1}^{f_2} S_\Phi (f) df.
\end{equation}
In experiments by Anton {\it et al.}, the mean-square flux noise in SQUIDs was found to increase with increasing temperature with $f_1=10^{-4} \text{ Hz}$ and $f_2=10^{9} \text{ Hz}$~\cite{Anton2013}.  In our simulations, the mean-square flux noise is equivalent to the total noise power with $f_1=0 \text{ MCS}^{-1}$ and $f_2=0.5 \text{ MCS}^{-1}$.  The total noise power is calculated for the isotropic ferromagnetic, noninteracting anisotropic, and isotropic antiferromagnetic Heisenberg systems for $0.5 \leq T \leq 10$.  The total noise power as a function of temperature is shown in Fig.~\ref{f_total-power}.  

In our simulations, only the Heisenberg antiferromagnet shows a total noise power that increases with temperature, though the curvature of the plot differs from that of Anton {\it et al.} \cite{Anton2013}. Furthermore, Anton {\it et al.} \cite{Anton2013} find that the mean square flux noise increases by two to three orders of magnitude as the temperature increases from 0.1 K to 4 K while our antiferromagnetic simulations find an increase of less than one order of magnitude.

\begin{figure}
\centering
\includegraphics[width=0.95 \linewidth]{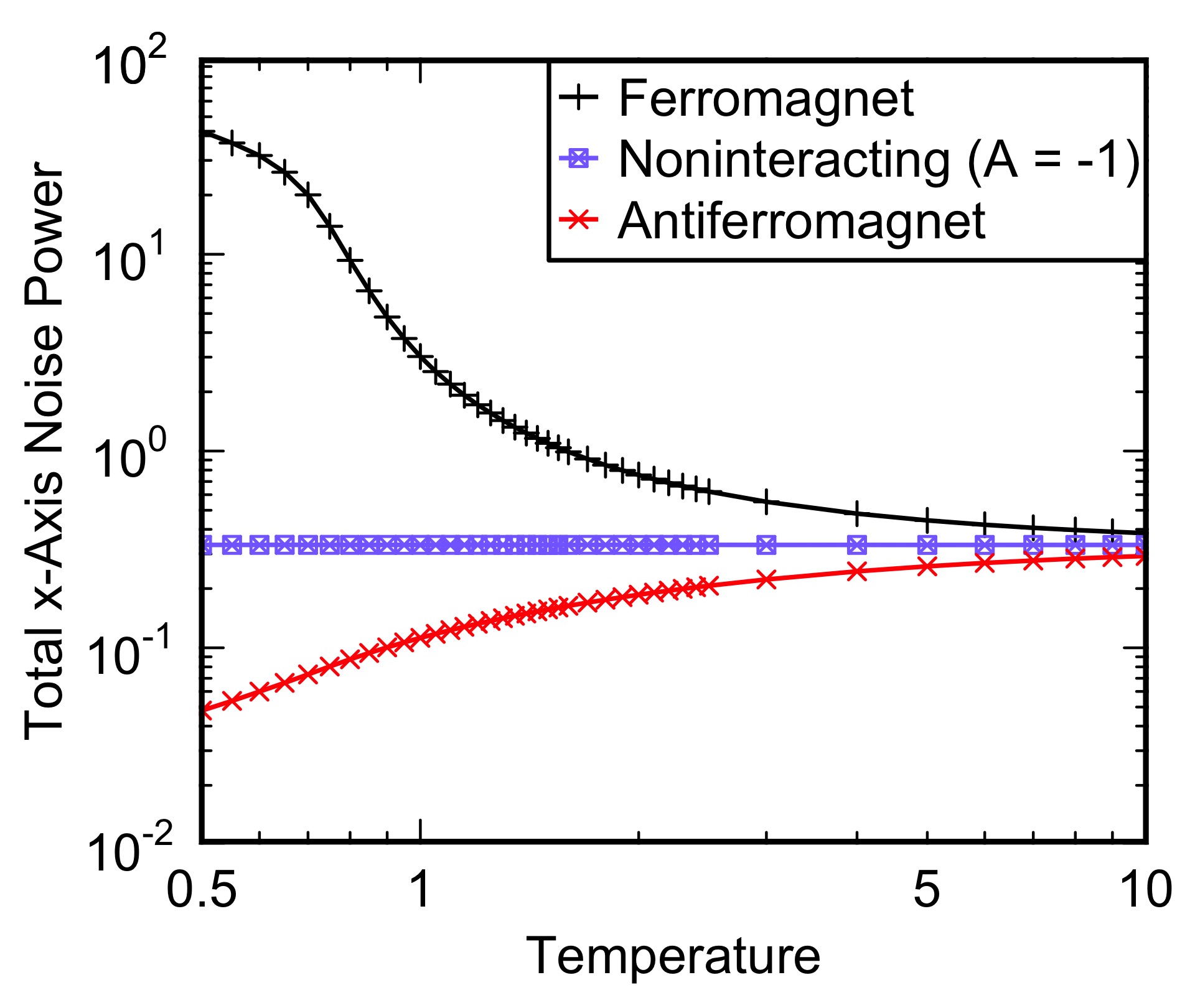}
\caption{Total noise power ($\int_{0 \text{ MCS}^{-1}}^{0.5 \text { MCS}^{-1}} S_x(f) df$) of the total magnetic moment versus temperature of the 2D isotropic ($A=0$) ferromagnetic ($J_{ij}=1$), noninteracting ($J_{ij}=0$) anisotropic ($A=-1$), and isotropic ($A=0$) antiferromagnetic ($J_{ij}=-1$) Heisenberg models for $0.5 \leq T \leq 10$.}
\label{f_total-power}
\end{figure}

\section{Summary and Discussion}
Monte Carlo simulations of various spin models with nearest-neighbor exchange were carried out on 2D lattices to determine which interactions yield $\alpha \sim 1$ at low temperatures as well as to find cases of spectral pivoting.  We find that the spin glass systems produce noise exponents that best match experiment~\cite{Wellstood1987,Anton2013,Kempf2016}.  In simulations, pivoting at high frequencies occurs as a result of the proximity of a low frequency knee and the aliasing of the noise power spectra.  In experiments, aliasing can be avoided by using low-pass filters~\cite{Clarke2004}.  This paper does not explain pivoting seen in experiments but it can explain the pivoting we see in some simulations.  We note that in our simulations pivoting is most evident at temperatures high compared with the magnetic transition temperature. Presumably this is consistent with experiments where there is no consistent evidence of a magnetic transition. 

This paper also does not explain why the mean-square flux noise increases with temperature in the experiment by Anton {\it et al.}~\cite{Anton2013}.  We find that this increase is characteristic of antiferromagnetic interactions between spins, for which there is no other experimental evidence. Furthermore, the increase in the mean square flux noise is 2 to 3 orders of magnitude in experiment compared to less than one order of magnitude in simulations. One way to interpret the experimental results would be to say that it implies an increase by a couple of orders of magnitude in the number of spins as the temperature increases; this is an interpretation that would be difficult to explain in the context of most spin models that have a fixed number of spins. Thus, the experimental findings imply that additional sources of flux noise become more prominent with increasing temperature. What these additional sources are is unknown.

\section{Acknowledgements}
We thank Herv{\'e} Carruzzo for helpful discussions.
RW acknowledges the support by U.S. Department of Energy, Office of Science, Basic Energy Sciences, under Award DE-FG02-05ER46237.
This work was performed in part at the Aspen Center for Physics, which is supported by National Science Foundation grant PHY-1607611.
\clearpage
\vfill

\appendix
\section{Fitting}
\label{a_fitting}
The process for fitting $A^2/f^\alpha$ to the noise power spectra to determine the noise exponent starts with dividing the noise power spectra into frequency segments as shown in Fig.~\ref{f_fitting-method}(a).  The segments  from $f_i$ to $f_{i+1}$ range from $i=1$ to $i=N_{\text{segments}}-1$.  The segments follow the condition $f_{i+1}/f_i=10^{0.1}$.  The function $A^2/f^\alpha$ is fit to each segment starting at the lowest frequency segment.

\begin{figure}
\centering
\includegraphics[width=0.95 \linewidth]{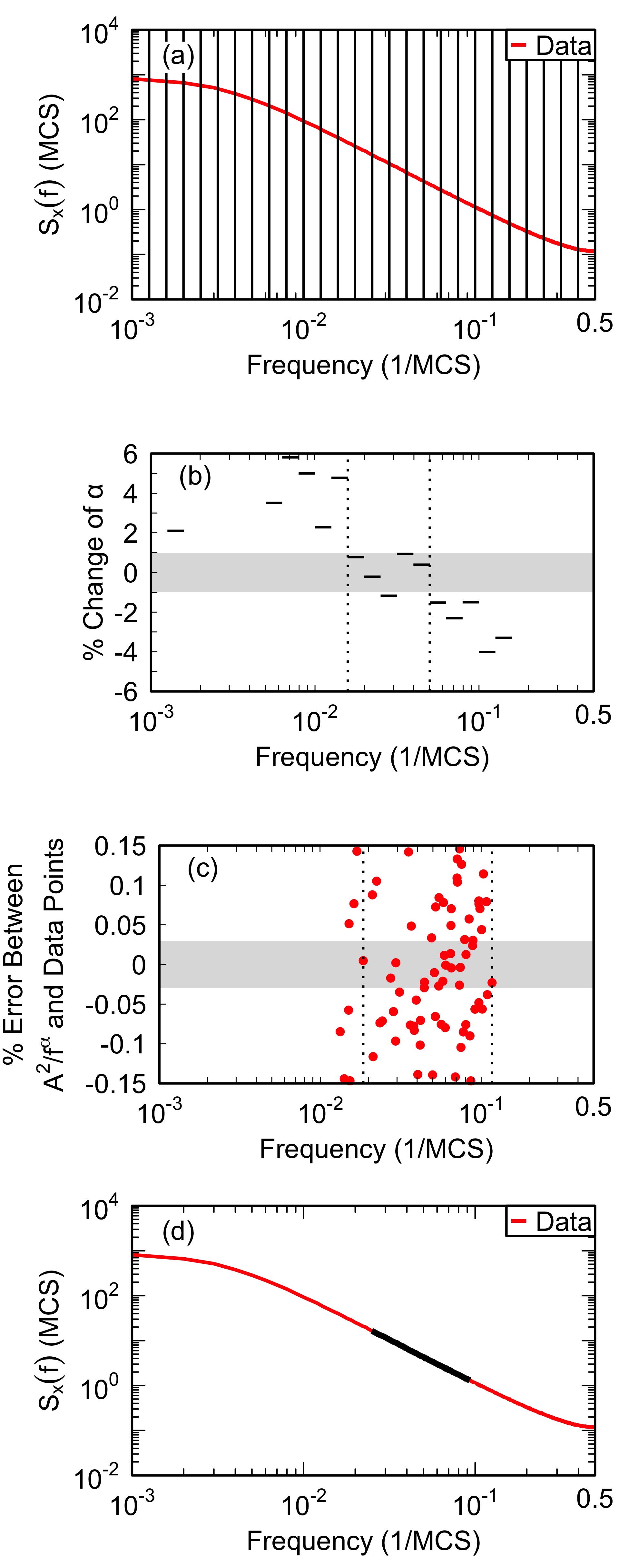}
\caption{Fitting method using the power spectrum averaged over 100 segments of the isotropic ($A=0$) Heisenberg ferromagnet ($J_{ij}$=1) at $T=0.9$.  (a) Segmented power spectra: noise power $S_x(f)$ of the x-component of the total magnetic moment versus frequency.  (b) Relative change in exponent versus frequency for the segmented power spectra. (c) Percent error between the power spectra fit and data versus frequency. (d) Noise power $S_x(f)$ of the x-component of the total magnetic moment versus frequency.  The data segment used for fitting is shown in black. }
\label{f_fitting-method}
\end{figure}

The relative change in the noise exponent between the current segment and the previous one is calculated.  An example of the relative change is shown in Fig.~\ref{f_fitting-method}(b).  If the relative change is less than $1\%$ and the noise exponent is greater than 0.2, then the frequency range of the segment is noted.  The lower and upper limits of the new frequency fit range are found by combining all of these segments.  In the figure, this is shown by two vertical dotted lines.  A fit is performed in this region.

The percent error of the fit is the percent difference between $A^2/f^\alpha$ evaluated at a particular $f$ and the noise power at $f$ from the data.  The percent error is evaluated for all frequency data points within the fitted frequency range and is shown in Fig.~\ref{f_fitting-method}(c).  A new frequency range is defined by the maximum and minimum frequencies corresponding to percent differences of less than $0.03\%$ as shown by vertical dotted lines.

This frequency range is shortened by increasing the lower limit by $30\%$ and decreasing the upper limit by $20\%$.  By reducing the frequency range, the fit region will not be within the knee and aliasing regions.  These percentages that were found by trial and error work well for all models presented.  The final fit of $A^2/f^\alpha$ is performed within this new frequency range.  The power spectra and final fit region of the power spectra are shown in Fig.~\ref{f_fitting-method}(d).

\section{High-Temperature Pivot of the Ising Spin Glass}
\label{a_ising-pivot}
As seen in Figs.~\ref{f_pivot-all} and \ref{f_pivot-ferromagnet} in Sec.~\ref{s_pivoting}, the noise power spectra tend to pivot at high temperature.  The noise power as a function of frequency at high temperature for the 2D Ising spin glass is shown in Fig.~\ref{f_pivot-ising-high}.  Although the power spectra pivots, the noise exponent is negative.  This is because Ising spin flips are always $180^\circ$ rotations.  At high temperature, Ising spins have a high probability of flipping at every time step which increases the noise at $f=0.5 \text{ MCS}^{-1}$.  In the infinite-temperature limit, the power spectrum would be a delta function peaked at $f=0.5 \text{ MCS}^{-1}$.
\begin{figure}
\centering
\includegraphics[width=\linewidth]{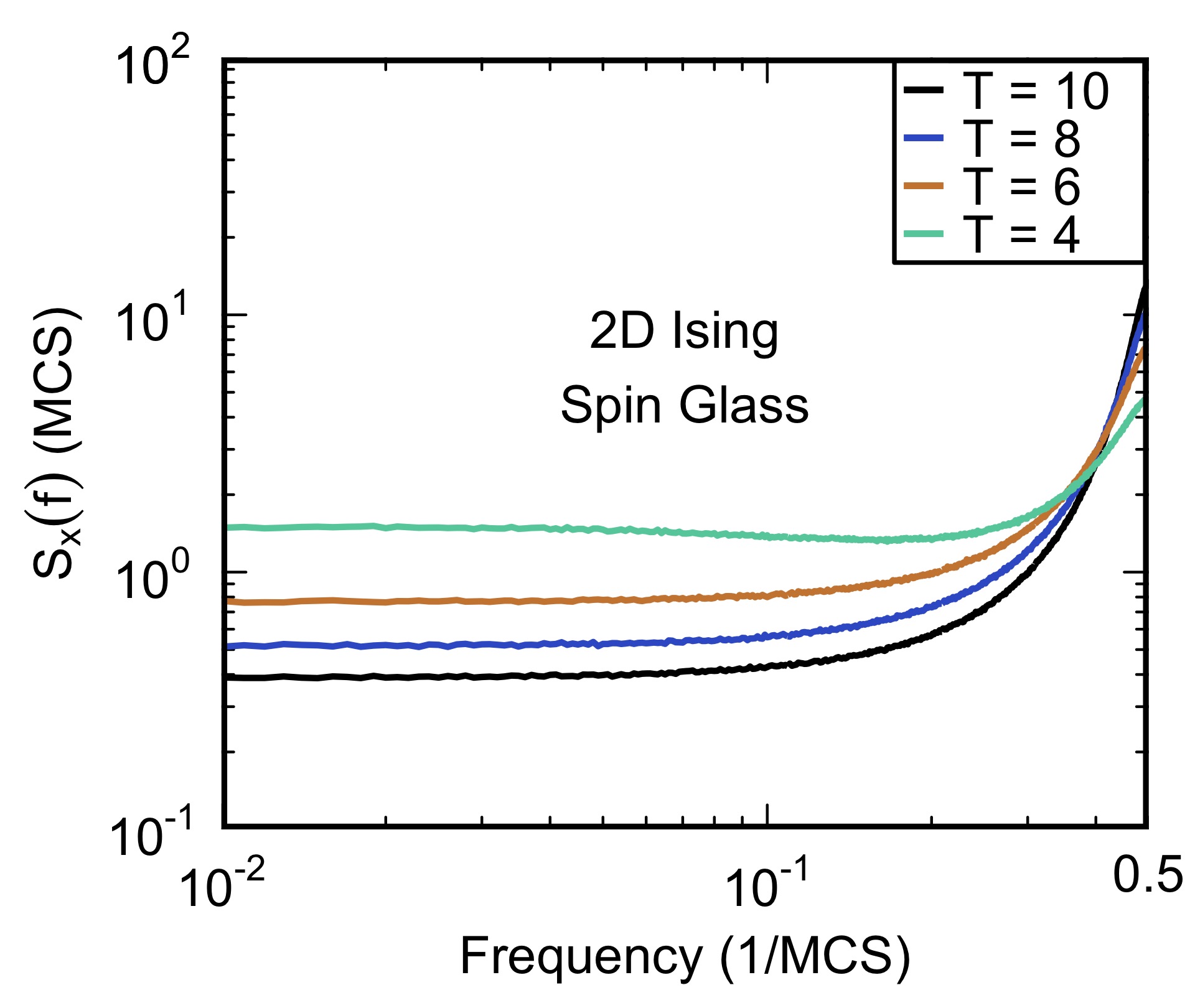}
\caption{Noise power $S(f)$ of the total magnetic moment of a 2D Ising spin glass ($\langle J_{ij} \rangle=0, \sigma_{J_{ij}}=1$) versus frequency.  Spectral pivoting of power spectra averaged over 100 segments for $4 \leq T \leq 10$.}
\label{f_pivot-ising-high}
\end{figure}

\section{Crossing Condition}
The experimental results of Anton {\it et al.} indicate that at high temperatures, the temperature dependence of the noise amplitude can be described by $A^2(T)=A^2_0T^\gamma$ and the noise exponent can be described by $\alpha(T)=\alpha_0 \ln(T)+\alpha_1$~\cite{Anton2013}.  Monte Carlo simulations also find that these relations hold at high temperature.   In addition, in both theory and experiment, the crossing frequency $f_c$ as a function of temperature does not change significantly.  This can be expressed as
\begin{equation}
\frac{dS(f,T)}{dT} \bigg|_{f=f_c}=0.
\end{equation}
Using $S(f,T)=A^2(T)/f^{\alpha(T)}$, we can relate the noise amplitudes to noise exponents:
\begin{equation}
\begin{aligned}
\frac{dS(f,T)}{dT} \bigg|_{f=f_c} &= \frac{d}{dT} \left( \frac{A^2(T)}{f^{\alpha(T)}}\right) \bigg|_{f=f_c}
 \\ &= \left(A^2(T)\frac{d}{dT}\frac{1}{f^{\alpha(T)}}+\frac{1}{f^{\alpha(T)}}\frac{d}{dT}A^2(T)\right) \bigg|_{f=f_c}
  \\ &= \left( A^2(T) \left( -\frac{\ln(f)}{f^{\alpha(T)}} \frac{d\alpha(T)}{dT} \right) + \frac{A^2_0}{f^{\alpha(T)}} T^{\gamma-1} \gamma \right) \bigg|_{f=f_c} 
  \\ &= -S(f_c) \ln(f_c) \frac{\alpha_0}{T}+\frac{S(f_c)}{T}\gamma
   \\ &=0.
\end{aligned}
\end{equation}
This gives the crossing condition
\begin{equation}
\gamma=\alpha_0 \ln(f_c).
\end{equation}

\clearpage

\begin{thebibliography}{29}%
\makeatletter
\providecommand \@ifxundefined [1]{%
 \@ifx{#1\undefined}
}%
\providecommand \@ifnum [1]{%
 \ifnum #1\expandafter \@firstoftwo
 \else \expandafter \@secondoftwo
 \fi
}%
\providecommand \@ifx [1]{%
 \ifx #1\expandafter \@firstoftwo
 \else \expandafter \@secondoftwo
 \fi
}%
\providecommand \natexlab [1]{#1}%
\providecommand \enquote  [1]{``#1''}%
\providecommand \bibnamefont  [1]{#1}%
\providecommand \bibfnamefont [1]{#1}%
\providecommand \citenamefont [1]{#1}%
\providecommand \href@noop [0]{\@secondoftwo}%
\providecommand \href [0]{\begingroup \@sanitize@url \@href}%
\providecommand \@href[1]{\@@startlink{#1}\@@href}%
\providecommand \@@href[1]{\endgroup#1\@@endlink}%
\providecommand \@sanitize@url [0]{\catcode `\\12\catcode `\$12\catcode
  `\&12\catcode `\#12\catcode `\^12\catcode `\_12\catcode `\%12\relax}%
\providecommand \@@startlink[1]{}%
\providecommand \@@endlink[0]{}%
\providecommand \url  [0]{\begingroup\@sanitize@url \@url }%
\providecommand \@url [1]{\endgroup\@href {#1}{\urlprefix }}%
\providecommand \urlprefix  [0]{URL }%
\providecommand \Eprint [0]{\href }%
\providecommand \doibase [0]{https://doi.org/}%
\providecommand \selectlanguage [0]{\@gobble}%
\providecommand \bibinfo  [0]{\@secondoftwo}%
\providecommand \bibfield  [0]{\@secondoftwo}%
\providecommand \translation [1]{[#1]}%
\providecommand \BibitemOpen [0]{}%
\providecommand \bibitemStop [0]{}%
\providecommand \bibitemNoStop [0]{.\EOS\space}%
\providecommand \EOS [0]{\spacefactor3000\relax}%
\providecommand \BibitemShut  [1]{\csname bibitem#1\endcsname}%
\let\auto@bib@innerbib\@empty
\bibitem [{\citenamefont {Bialczak}\ \emph {et~al.}(2007)\citenamefont
  {Bialczak}, \citenamefont {McDermott}, \citenamefont {Ansmann}, \citenamefont
  {Hofheinz}, \citenamefont {Katz}, \citenamefont {Lucero}, \citenamefont
  {Neeley}, \citenamefont {O'Connell}, \citenamefont {Wang}, \citenamefont
  {Cleland},\ and\ \citenamefont {Martinis}}]{Bialczak2007}%
  \BibitemOpen
  \bibfield  {author} {\bibinfo {author} {\bibfnamefont {R.~C.}\ \bibnamefont
  {Bialczak}}, \bibinfo {author} {\bibfnamefont {R.}~\bibnamefont {McDermott}},
  \bibinfo {author} {\bibfnamefont {M.}~\bibnamefont {Ansmann}}, \bibinfo
  {author} {\bibfnamefont {M.}~\bibnamefont {Hofheinz}}, \bibinfo {author}
  {\bibfnamefont {N.}~\bibnamefont {Katz}}, \bibinfo {author} {\bibfnamefont
  {E.}~\bibnamefont {Lucero}}, \bibinfo {author} {\bibfnamefont
  {M.}~\bibnamefont {Neeley}}, \bibinfo {author} {\bibfnamefont {A.~D.}\
  \bibnamefont {O'Connell}}, \bibinfo {author} {\bibfnamefont {H.}~\bibnamefont
  {Wang}}, \bibinfo {author} {\bibfnamefont {A.~N.}\ \bibnamefont {Cleland}},\
  and\ \bibinfo {author} {\bibfnamefont {J.~M.}\ \bibnamefont {Martinis}},\
  }\bibfield  {title} {\bibinfo {title} {$1/f$ flux noise in josephson phase
  qubits},\ }\href {https://doi.org/10.1103/PhysRevLett.99.187006} {\bibfield
  {journal} {\bibinfo  {journal} {Phys. Rev. Lett.}\ }\textbf {\bibinfo
  {volume} {99}},\ \bibinfo {pages} {187006} (\bibinfo {year}
  {2007})}\BibitemShut {NoStop}%
\bibitem [{\citenamefont {Wellstood}\ \emph {et~al.}(1987)\citenamefont
  {Wellstood}, \citenamefont {Urbina},\ and\ \citenamefont
  {Clarke}}]{Wellstood1987}%
  \BibitemOpen
  \bibfield  {author} {\bibinfo {author} {\bibfnamefont {F.~C.}\ \bibnamefont
  {Wellstood}}, \bibinfo {author} {\bibfnamefont {C.}~\bibnamefont {Urbina}},\
  and\ \bibinfo {author} {\bibfnamefont {J.}~\bibnamefont {Clarke}},\
  }\bibfield  {title} {\bibinfo {title} {Low‐frequency noise in dc
  superconducting quantum interference devices below 1 k},\ }\href
  {https://doi.org/10.1063/1.98041} {\bibfield  {journal} {\bibinfo  {journal}
  {Applied Physics Letters}\ }\textbf {\bibinfo {volume} {50}},\ \bibinfo
  {pages} {772} (\bibinfo {year} {1987})}\BibitemShut {NoStop}%
\bibitem [{\citenamefont {Anton}\ \emph {et~al.}(2013)\citenamefont {Anton},
  \citenamefont {Birenbaum}, \citenamefont {O'Kelley}, \citenamefont
  {Bolkhovsky}, \citenamefont {Braje}, \citenamefont {Fitch}, \citenamefont
  {Neeley}, \citenamefont {Hilton}, \citenamefont {Cho}, \citenamefont {Irwin},
  \citenamefont {Wellstood}, \citenamefont {Oliver}, \citenamefont {Shnirman},\
  and\ \citenamefont {Clarke}}]{Anton2013}%
  \BibitemOpen
  \bibfield  {author} {\bibinfo {author} {\bibfnamefont {S.~M.}\ \bibnamefont
  {Anton}}, \bibinfo {author} {\bibfnamefont {J.~S.}\ \bibnamefont
  {Birenbaum}}, \bibinfo {author} {\bibfnamefont {S.~R.}\ \bibnamefont
  {O'Kelley}}, \bibinfo {author} {\bibfnamefont {V.}~\bibnamefont
  {Bolkhovsky}}, \bibinfo {author} {\bibfnamefont {D.~A.}\ \bibnamefont
  {Braje}}, \bibinfo {author} {\bibfnamefont {G.}~\bibnamefont {Fitch}},
  \bibinfo {author} {\bibfnamefont {M.}~\bibnamefont {Neeley}}, \bibinfo
  {author} {\bibfnamefont {G.~C.}\ \bibnamefont {Hilton}}, \bibinfo {author}
  {\bibfnamefont {H.-M.}\ \bibnamefont {Cho}}, \bibinfo {author} {\bibfnamefont
  {K.~D.}\ \bibnamefont {Irwin}}, \bibinfo {author} {\bibfnamefont {F.~C.}\
  \bibnamefont {Wellstood}}, \bibinfo {author} {\bibfnamefont {W.~D.}\
  \bibnamefont {Oliver}}, \bibinfo {author} {\bibfnamefont {A.}~\bibnamefont
  {Shnirman}},\ and\ \bibinfo {author} {\bibfnamefont {J.}~\bibnamefont
  {Clarke}},\ }\bibfield  {title} {\bibinfo {title} {Magnetic flux noise in dc
  squids: Temperature and geometry dependence},\ }\href
  {https://doi.org/10.1103/PhysRevLett.110.147002} {\bibfield  {journal}
  {\bibinfo  {journal} {Phys. Rev. Lett.}\ }\textbf {\bibinfo {volume} {110}},\
  \bibinfo {pages} {147002} (\bibinfo {year} {2013})}\BibitemShut {NoStop}%
\bibitem [{\citenamefont {Kempf}\ \emph {et~al.}(2016)\citenamefont {Kempf},
  \citenamefont {Ferring},\ and\ \citenamefont {Enss}}]{Kempf2016}%
  \BibitemOpen
  \bibfield  {author} {\bibinfo {author} {\bibfnamefont {S.}~\bibnamefont
  {Kempf}}, \bibinfo {author} {\bibfnamefont {A.}~\bibnamefont {Ferring}},\
  and\ \bibinfo {author} {\bibfnamefont {C.}~\bibnamefont {Enss}},\ }\bibfield
  {title} {\bibinfo {title} {Towards noise engineering: Recent insights in
  low-frequency excess flux noise of superconducting quantum devices},\ }\href
  {https://doi.org/10.1063/1.4965293} {\bibfield  {journal} {\bibinfo
  {journal} {Appl. Phys. Lett.}\ }\textbf {\bibinfo {volume} {109}},\ \bibinfo
  {pages} {162601} (\bibinfo {year} {2016})}\BibitemShut {NoStop}%
\bibitem [{\citenamefont {Bluhm}\ \emph {et~al.}(2009)\citenamefont {Bluhm},
  \citenamefont {Bert}, \citenamefont {Koshnick}, \citenamefont {Huber},\ and\
  \citenamefont {Moler}}]{Bluhm2009}%
  \BibitemOpen
  \bibfield  {author} {\bibinfo {author} {\bibfnamefont {H.}~\bibnamefont
  {Bluhm}}, \bibinfo {author} {\bibfnamefont {J.~A.}\ \bibnamefont {Bert}},
  \bibinfo {author} {\bibfnamefont {N.~C.}\ \bibnamefont {Koshnick}}, \bibinfo
  {author} {\bibfnamefont {M.~E.}\ \bibnamefont {Huber}},\ and\ \bibinfo
  {author} {\bibfnamefont {K.~A.}\ \bibnamefont {Moler}},\ }\bibfield  {title}
  {\bibinfo {title} {Spinlike susceptibility of metallic and insulating thin
  films at low temperature},\ }\href
  {https://doi.org/10.1103/PhysRevLett.103.026805} {\bibfield  {journal}
  {\bibinfo  {journal} {Phys. Rev. Lett.}\ }\textbf {\bibinfo {volume} {103}},\
  \bibinfo {pages} {026805} (\bibinfo {year} {2009})}\BibitemShut {NoStop}%
\bibitem [{\citenamefont {Sendelbach}\ \emph {et~al.}(2008)\citenamefont
  {Sendelbach}, \citenamefont {Hover}, \citenamefont {Kittel}, \citenamefont
  {M\"uck}, \citenamefont {Martinis},\ and\ \citenamefont
  {McDermott}}]{Sendelbach2008}%
  \BibitemOpen
  \bibfield  {author} {\bibinfo {author} {\bibfnamefont {S.}~\bibnamefont
  {Sendelbach}}, \bibinfo {author} {\bibfnamefont {D.}~\bibnamefont {Hover}},
  \bibinfo {author} {\bibfnamefont {A.}~\bibnamefont {Kittel}}, \bibinfo
  {author} {\bibfnamefont {M.}~\bibnamefont {M\"uck}}, \bibinfo {author}
  {\bibfnamefont {J.~M.}\ \bibnamefont {Martinis}},\ and\ \bibinfo {author}
  {\bibfnamefont {R.}~\bibnamefont {McDermott}},\ }\bibfield  {title} {\bibinfo
  {title} {Magnetism in squids at millikelvin temperatures},\ }\href
  {https://doi.org/10.1103/PhysRevLett.100.227006} {\bibfield  {journal}
  {\bibinfo  {journal} {Phys. Rev. Lett.}\ }\textbf {\bibinfo {volume} {100}},\
  \bibinfo {pages} {227006} (\bibinfo {year} {2008})}\BibitemShut {NoStop}%
\bibitem [{\citenamefont {Sendelbach}\ \emph {et~al.}(2009)\citenamefont
  {Sendelbach}, \citenamefont {Hover}, \citenamefont {M\"uck},\ and\
  \citenamefont {McDermott}}]{Sendelbach2009}%
  \BibitemOpen
  \bibfield  {author} {\bibinfo {author} {\bibfnamefont {S.}~\bibnamefont
  {Sendelbach}}, \bibinfo {author} {\bibfnamefont {D.}~\bibnamefont {Hover}},
  \bibinfo {author} {\bibfnamefont {M.}~\bibnamefont {M\"uck}},\ and\ \bibinfo
  {author} {\bibfnamefont {R.}~\bibnamefont {McDermott}},\ }\bibfield  {title}
  {\bibinfo {title} {Complex inductance, excess noise, and surface magnetism in
  dc squids},\ }\href {https://doi.org/10.1103/PhysRevLett.103.117001}
  {\bibfield  {journal} {\bibinfo  {journal} {Phys. Rev. Lett.}\ }\textbf
  {\bibinfo {volume} {103}},\ \bibinfo {pages} {117001} (\bibinfo {year}
  {2009})}\BibitemShut {NoStop}%
\bibitem [{\citenamefont {Wang}\ \emph {et~al.}(2015)\citenamefont {Wang},
  \citenamefont {Shi}, \citenamefont {Hu}, \citenamefont {Han}, \citenamefont
  {Yu},\ and\ \citenamefont {Wu}}]{Wang2015}%
  \BibitemOpen
  \bibfield  {author} {\bibinfo {author} {\bibfnamefont {H.}~\bibnamefont
  {Wang}}, \bibinfo {author} {\bibfnamefont {C.}~\bibnamefont {Shi}}, \bibinfo
  {author} {\bibfnamefont {J.}~\bibnamefont {Hu}}, \bibinfo {author}
  {\bibfnamefont {S.}~\bibnamefont {Han}}, \bibinfo {author} {\bibfnamefont
  {C.~C.}\ \bibnamefont {Yu}},\ and\ \bibinfo {author} {\bibfnamefont {R.~Q.}\
  \bibnamefont {Wu}},\ }\bibfield  {title} {\bibinfo {title} {Candidate source
  of flux noise in squids: Adsorbed oxygen molecules},\ }\href
  {https://doi.org/10.1103/PhysRevLett.115.077002} {\bibfield  {journal}
  {\bibinfo  {journal} {Phys. Rev. Lett.}\ }\textbf {\bibinfo {volume} {115}},\
  \bibinfo {pages} {077002} (\bibinfo {year} {2015})}\BibitemShut {NoStop}%
\bibitem [{\citenamefont {Kumar}\ \emph {et~al.}(2016)\citenamefont {Kumar},
  \citenamefont {Sendelbach}, \citenamefont {Beck}, \citenamefont {Freeland},
  \citenamefont {Wang}, \citenamefont {Wang}, \citenamefont {Yu}, \citenamefont
  {Wu}, \citenamefont {Pappas},\ and\ \citenamefont {McDermott}}]{Kumar2016}%
  \BibitemOpen
  \bibfield  {author} {\bibinfo {author} {\bibfnamefont {P.}~\bibnamefont
  {Kumar}}, \bibinfo {author} {\bibfnamefont {S.}~\bibnamefont {Sendelbach}},
  \bibinfo {author} {\bibfnamefont {M.~A.}\ \bibnamefont {Beck}}, \bibinfo
  {author} {\bibfnamefont {J.~W.}\ \bibnamefont {Freeland}}, \bibinfo {author}
  {\bibfnamefont {Z.}~\bibnamefont {Wang}}, \bibinfo {author} {\bibfnamefont
  {H.}~\bibnamefont {Wang}}, \bibinfo {author} {\bibfnamefont {C.~C.}\
  \bibnamefont {Yu}}, \bibinfo {author} {\bibfnamefont {R.~Q.}\ \bibnamefont
  {Wu}}, \bibinfo {author} {\bibfnamefont {D.~P.}\ \bibnamefont {Pappas}},\
  and\ \bibinfo {author} {\bibfnamefont {R.}~\bibnamefont {McDermott}},\
  }\bibfield  {title} {\bibinfo {title} {Origin and reduction of $1/f$ magnetic
  flux noise in superconducting devices},\ }\href
  {https://doi.org/10.1103/PhysRevApplied.6.041001} {\bibfield  {journal}
  {\bibinfo  {journal} {Phys. Rev. Applied}\ }\textbf {\bibinfo {volume} {6}},\
  \bibinfo {pages} {041001(R)} (\bibinfo {year} {2016})}\BibitemShut {NoStop}%
\bibitem [{\citenamefont {Dutta}\ and\ \citenamefont {Horn}(1981)}]{Dutta1981}%
  \BibitemOpen
  \bibfield  {author} {\bibinfo {author} {\bibfnamefont {P.}~\bibnamefont
  {Dutta}}\ and\ \bibinfo {author} {\bibfnamefont {P.~M.}\ \bibnamefont
  {Horn}},\ }\bibfield  {title} {\bibinfo {title} {Low-frequency fluctuations
  in solids: $\frac{1}{f}$ noise},\ }\href
  {https://doi.org/10.1103/RevModPhys.53.497} {\bibfield  {journal} {\bibinfo
  {journal} {Rev. Mod. Phys.}\ }\textbf {\bibinfo {volume} {53}},\ \bibinfo
  {pages} {497} (\bibinfo {year} {1981})}\BibitemShut {NoStop}%
\bibitem [{\citenamefont {Koch}\ \emph {et~al.}(2007)\citenamefont {Koch},
  \citenamefont {DiVincenzo},\ and\ \citenamefont {Clarke}}]{Koch2007}%
  \BibitemOpen
  \bibfield  {author} {\bibinfo {author} {\bibfnamefont {R.~H.}\ \bibnamefont
  {Koch}}, \bibinfo {author} {\bibfnamefont {D.~P.}\ \bibnamefont
  {DiVincenzo}},\ and\ \bibinfo {author} {\bibfnamefont {J.}~\bibnamefont
  {Clarke}},\ }\bibfield  {title} {\bibinfo {title} {Model for $1/f$ flux noise
  in squids and qubits},\ }\href
  {https://doi.org/10.1103/PhysRevLett.98.267003} {\bibfield  {journal}
  {\bibinfo  {journal} {Phys. Rev. Lett.}\ }\textbf {\bibinfo {volume} {98}},\
  \bibinfo {pages} {267003} (\bibinfo {year} {2007})}\BibitemShut {NoStop}%
\bibitem [{\citenamefont {Choi}\ \emph {et~al.}(2009)\citenamefont {Choi},
  \citenamefont {Lee}, \citenamefont {Louie},\ and\ \citenamefont
  {Clarke}}]{Choi2009}%
  \BibitemOpen
  \bibfield  {author} {\bibinfo {author} {\bibfnamefont {S.~K.}~\bibnamefont
  {Choi}}, \bibinfo {author} {\bibfnamefont {D.-H.}\ \bibnamefont {Lee}},
  \bibinfo {author} {\bibfnamefont {S.~G.}\ \bibnamefont {Louie}},\ and\
  \bibinfo {author} {\bibfnamefont {J.}~\bibnamefont {Clarke}},\ }\bibfield
  {title} {\bibinfo {title} {Localization of metal-induced gap states at the
  metal-insulator interface: Origin of flux noise in squids and superconducting
  qubits},\ }\href {https://doi.org/10.1103/PhysRevLett.103.197001} {\bibfield
  {journal} {\bibinfo  {journal} {Phys. Rev. Lett.}\ }\textbf {\bibinfo
  {volume} {103}},\ \bibinfo {pages} {197001} (\bibinfo {year}
  {2009})}\BibitemShut {NoStop}%
\bibitem [{\citenamefont {de~Sousa}(2007)}]{deSousa2007}%
  \BibitemOpen
  \bibfield  {author} {\bibinfo {author} {\bibfnamefont {R.}~\bibnamefont
  {de~Sousa}},\ }\bibfield  {title} {\bibinfo {title} {Dangling-bond spin
  relaxation and magnetic $1/f$ noise from the amorphous-semiconductor/oxide
  interface: Theory},\ }\href {https://doi.org/10.1103/PhysRevB.76.245306}
  {\bibfield  {journal} {\bibinfo  {journal} {Phys. Rev. B}\ }\textbf {\bibinfo
  {volume} {76}},\ \bibinfo {pages} {245306} (\bibinfo {year}
  {2007})}\BibitemShut {NoStop}%
\bibitem [{\citenamefont {Faoro}\ and\ \citenamefont
  {Ioffe}(2008)}]{Faoro2008}%
  \BibitemOpen
  \bibfield  {author} {\bibinfo {author} {\bibfnamefont {L.}~\bibnamefont
  {Faoro}}\ and\ \bibinfo {author} {\bibfnamefont {L.~B.}\ \bibnamefont
  {Ioffe}},\ }\bibfield  {title} {\bibinfo {title} {Microscopic origin of
  low-frequency flux noise in josephson circuits},\ }\href
  {https://doi.org/10.1103/PhysRevLett.100.227005} {\bibfield  {journal}
  {\bibinfo  {journal} {Phys. Rev. Lett.}\ }\textbf {\bibinfo {volume} {100}},\
  \bibinfo {pages} {227005} (\bibinfo {year} {2008})}\BibitemShut {NoStop}%
\bibitem [{\citenamefont {Chen}\ and\ \citenamefont {Yu}(2010)}]{Chen2010}%
  \BibitemOpen
  \bibfield  {author} {\bibinfo {author} {\bibfnamefont {Z.}~\bibnamefont
  {Chen}}\ and\ \bibinfo {author} {\bibfnamefont {C.~C.}\ \bibnamefont {Yu}},\
  }\bibfield  {title} {\bibinfo {title} {Comparison of ising spin glass noise
  to flux and inductance noise in squids},\ }\href
  {https://doi.org/10.1103/PhysRevLett.104.247204} {\bibfield  {journal}
  {\bibinfo  {journal} {Phys. Rev. Lett.}\ }\textbf {\bibinfo {volume} {104}},\
  \bibinfo {pages} {247204} (\bibinfo {year} {2010})}\BibitemShut {NoStop}%
\bibitem [{\citenamefont {De}(2014)}]{De2014}%
  \BibitemOpen
  \bibfield  {author} {\bibinfo {author} {\bibfnamefont {A.}~\bibnamefont
  {De}},\ }\bibfield  {title} {\bibinfo {title} {Ising-glauber spin cluster
  model for temperature-dependent magnetization noise in squids},\ }\href
  {https://doi.org/10.1103/PhysRevLett.113.217002} {\bibfield  {journal}
  {\bibinfo  {journal} {Phys. Rev. Lett.}\ }\textbf {\bibinfo {volume} {113}},\
  \bibinfo {pages} {217002} (\bibinfo {year} {2014})}\BibitemShut {NoStop}%
\bibitem [{\citenamefont {De}(2019)}]{De2019}%
  \BibitemOpen
  \bibfield  {author} {\bibinfo {author} {\bibfnamefont {A.}~\bibnamefont
  {De}},\ }\bibfield  {title} {\bibinfo {title} {$1/f$ flux noise in
  low-${T}_{c}$ squids due to superparamagnetic phase transitions in defect
  clusters},\ }\href {https://doi.org/10.1103/PhysRevB.99.024305} {\bibfield
  {journal} {\bibinfo  {journal} {Phys. Rev. B}\ }\textbf {\bibinfo {volume}
  {99}},\ \bibinfo {pages} {024305} (\bibinfo {year} {2019})}\BibitemShut
  {NoStop}%
\bibitem [{\citenamefont {Lanting}\ \emph {et~al.}(2014)\citenamefont
  {Lanting}, \citenamefont {Amin}, \citenamefont {Berkley}, \citenamefont
  {Rich}, \citenamefont {Chen}, \citenamefont {LaForest},\ and\ \citenamefont
  {de~Sousa}}]{Lanting2014}%
  \BibitemOpen
  \bibfield  {author} {\bibinfo {author} {\bibfnamefont {T.}~\bibnamefont
  {Lanting}}, \bibinfo {author} {\bibfnamefont {M.~H.}\ \bibnamefont {Amin}},
  \bibinfo {author} {\bibfnamefont {A.~J.}\ \bibnamefont {Berkley}}, \bibinfo
  {author} {\bibfnamefont {C.}~\bibnamefont {Rich}}, \bibinfo {author}
  {\bibfnamefont {S.-F.}\ \bibnamefont {Chen}}, \bibinfo {author}
  {\bibfnamefont {S.}~\bibnamefont {LaForest}},\ and\ \bibinfo {author}
  {\bibfnamefont {R.}~\bibnamefont {de~Sousa}},\ }\bibfield  {title} {\bibinfo
  {title} {Evidence for temperature-dependent spin diffusion as a mechanism of
  intrinsic flux noise in squids},\ }\href
  {https://doi.org/10.1103/PhysRevB.89.014503} {\bibfield  {journal} {\bibinfo
  {journal} {Phys. Rev. B}\ }\textbf {\bibinfo {volume} {89}},\ \bibinfo
  {pages} {014503} (\bibinfo {year} {2014})}\BibitemShut {NoStop}%
\bibitem [{\citenamefont {Davis}\ and\ \citenamefont
  {Chamberlin}(2018)}]{Davis2018}%
  \BibitemOpen
  \bibfield  {author} {\bibinfo {author} {\bibfnamefont {B.~F.}\ \bibnamefont
  {Davis}}\ and\ \bibinfo {author} {\bibfnamefont {R.~V.}\ \bibnamefont
  {Chamberlin}},\ }\bibfield  {title} {\bibinfo {title} {1/f noise from a
  finite entropy bath: comparison with flux noise in squids},\ }\href
  {http://stacks.iop.org/1742-5468/2018/i=10/a=103206} {\bibfield  {journal}
  {\bibinfo  {journal} {Journal of Statistical Mechanics: Theory and
  Experiment}\ }\textbf {\bibinfo {volume} {2018}},\ \bibinfo {pages} {103206}
  (\bibinfo {year} {2018})}\BibitemShut {NoStop}%
\bibitem [{\citenamefont {Press}\ \emph {et~al.}(2007)\citenamefont {Press},
  \citenamefont {Teukolsky}, \citenamefont {Vetterling},\ and\ \citenamefont
  {Flannery}}]{Press2007}%
  \BibitemOpen
  \bibfield  {author} {\bibinfo {author} {\bibfnamefont {W.~H.}\ \bibnamefont
  {Press}}, \bibinfo {author} {\bibfnamefont {S.~A.}\ \bibnamefont
  {Teukolsky}}, \bibinfo {author} {\bibfnamefont {W.~T.}\ \bibnamefont
  {Vetterling}},\ and\ \bibinfo {author} {\bibfnamefont {B.~P.}\ \bibnamefont
  {Flannery}},\ }\href@noop {} {\emph {\bibinfo {title} {Numerical Recipes in
  C: The Art of Scientific Computing}}},\ \bibinfo {edition} {3rd}\ ed.\
  (\bibinfo  {publisher} {Cambridge University Press},\ \bibinfo {address} {New
  York},\ \bibinfo {year} {2007})\BibitemShut {NoStop}%
\bibitem [{\citenamefont {Harris}\ \emph {et~al.}(1973)\citenamefont {Harris},
  \citenamefont {Plischke},\ and\ \citenamefont {Zuckermann}}]{Harris1973}%
  \BibitemOpen
  \bibfield  {author} {\bibinfo {author} {\bibfnamefont {R.}~\bibnamefont
  {Harris}}, \bibinfo {author} {\bibfnamefont {M.}~\bibnamefont {Plischke}},\
  and\ \bibinfo {author} {\bibfnamefont {M.~J.}\ \bibnamefont {Zuckermann}},\
  }\bibfield  {title} {\bibinfo {title} {New model for amorphous magnetism},\
  }\href {https://doi.org/10.1103/PhysRevLett.31.160} {\bibfield  {journal}
  {\bibinfo  {journal} {Phys. Rev. Lett.}\ }\textbf {\bibinfo {volume} {31}},\
  \bibinfo {pages} {160} (\bibinfo {year} {1973})}\BibitemShut {NoStop}%
\bibitem [{\citenamefont {Brown}(1963)}]{Brown1963}%
  \BibitemOpen
  \bibfield  {author} {\bibinfo {author} {\bibfnamefont {W.~F.}\ \bibnamefont
  {Brown}},\ }\bibfield  {title} {\bibinfo {title} {Thermal fluctuations of a
  single-domain particle},\ }\href {https://doi.org/10.1103/PhysRev.130.1677}
  {\bibfield  {journal} {\bibinfo  {journal} {Phys. Rev.}\ }\textbf {\bibinfo
  {volume} {130}},\ \bibinfo {pages} {1677} (\bibinfo {year}
  {1963})}\BibitemShut {NoStop}%
\bibitem [{\citenamefont {Metropolis}\ \emph {et~al.}(1953)\citenamefont
  {Metropolis}, \citenamefont {Rosenbluth}, \citenamefont {Rosenbluth},
  \citenamefont {Teller},\ and\ \citenamefont {Teller}}]{Metropolis1953}%
  \BibitemOpen
  \bibfield  {author} {\bibinfo {author} {\bibfnamefont {N.}~\bibnamefont
  {Metropolis}}, \bibinfo {author} {\bibfnamefont {A.~W.}\ \bibnamefont
  {Rosenbluth}}, \bibinfo {author} {\bibfnamefont {M.~N.}\ \bibnamefont
  {Rosenbluth}}, \bibinfo {author} {\bibfnamefont {A.~H.}\ \bibnamefont
  {Teller}},\ and\ \bibinfo {author} {\bibfnamefont {E.}~\bibnamefont
  {Teller}},\ }\bibfield  {title} {\bibinfo {title} {Equation of state
  calculations by fast computing machines},\ }\href
  {https://doi.org/10.1063/1.1699114} {\bibfield  {journal} {\bibinfo
  {journal} {J. Chem. Phys.}\ }\textbf {\bibinfo {volume} {21}},\ \bibinfo
  {pages} {1087} (\bibinfo {year} {1953})}\BibitemShut {NoStop}%
\bibitem [{\citenamefont {Bhatt}\ and\ \citenamefont
  {Young}(1988)}]{Bhatt1988}%
  \BibitemOpen
  \bibfield  {author} {\bibinfo {author} {\bibfnamefont {R.~N.}\ \bibnamefont
  {Bhatt}}\ and\ \bibinfo {author} {\bibfnamefont {A.~P.}\ \bibnamefont
  {Young}},\ }\bibfield  {title} {\bibinfo {title} {Numerical studies of ising
  spin glasses in two, three, and four dimensions},\ }\href
  {https://doi.org/10.1103/PhysRevB.37.5606} {\bibfield  {journal} {\bibinfo
  {journal} {Phys. Rev. B}\ }\textbf {\bibinfo {volume} {37}},\ \bibinfo
  {pages} {5606} (\bibinfo {year} {1988})}\BibitemShut {NoStop}%
\bibitem [{\citenamefont {Frigo}\ and\ \citenamefont
  {Johnson}(2005)}]{Frigo2005}%
  \BibitemOpen
  \bibfield  {author} {\bibinfo {author} {\bibfnamefont {M.}~\bibnamefont
  {Frigo}}\ and\ \bibinfo {author} {\bibfnamefont {S.~G.}\ \bibnamefont
  {Johnson}},\ }\bibfield  {title} {\bibinfo {title} {The design and
  implementation of {FFTW3}},\ }\href@noop {} {\bibfield  {journal} {\bibinfo
  {journal} {Proceedings of the IEEE}\ }\textbf {\bibinfo {volume} {93}},\
  \bibinfo {pages} {216} (\bibinfo {year} {2005})},\ \bibinfo {note} {special
  issue on ``Program Generation, Optimization, and Platform
  Adaptation''}\BibitemShut {NoStop}%
\bibitem [{\citenamefont {Chen}\ and\ \citenamefont {Yu}(2007)}]{Chen2007}%
  \BibitemOpen
  \bibfield  {author} {\bibinfo {author} {\bibfnamefont {Z.}~\bibnamefont
  {Chen}}\ and\ \bibinfo {author} {\bibfnamefont {C.~C.}\ \bibnamefont {Yu}},\
  }\bibfield  {title} {\bibinfo {title} {Measurement-noise maximum as a
  signature of a phase transition},\ }\href
  {https://doi.org/10.1103/PhysRevLett.98.057204} {\bibfield  {journal}
  {\bibinfo  {journal} {Phys. Rev. Lett.}\ }\textbf {\bibinfo {volume} {98}},\
  \bibinfo {pages} {057204} (\bibinfo {year} {2007})}\BibitemShut {NoStop}%
\bibitem [{\citenamefont {Mermin}\ and\ \citenamefont
  {Wagner}(1966)}]{Mermin1966}%
  \BibitemOpen
  \bibfield  {author} {\bibinfo {author} {\bibfnamefont {N.~D.}\ \bibnamefont
  {Mermin}}\ and\ \bibinfo {author} {\bibfnamefont {H.}~\bibnamefont
  {Wagner}},\ }\bibfield  {title} {\bibinfo {title} {Absence of ferromagnetism
  or antiferromagnetism in one- or two-dimensional isotropic heisenberg
  models},\ }\href {https://doi.org/10.1103/PhysRevLett.17.1133} {\bibfield
  {journal} {\bibinfo  {journal} {Phys. Rev. Lett.}\ }\textbf {\bibinfo
  {volume} {17}},\ \bibinfo {pages} {1133} (\bibinfo {year}
  {1966})}\BibitemShut {NoStop}%
\bibitem [{\citenamefont {Hohenberg}(1967)}]{Hohenberg1967}%
  \BibitemOpen
  \bibfield  {author} {\bibinfo {author} {\bibfnamefont {P.~C.}\ \bibnamefont
  {Hohenberg}},\ }\bibfield  {title} {\bibinfo {title} {Existence of long-range
  order in one and two dimensions},\ }\href
  {https://doi.org/10.1103/PhysRev.158.383} {\bibfield  {journal} {\bibinfo
  {journal} {Phys. Rev.}\ }\textbf {\bibinfo {volume} {158}},\ \bibinfo {pages}
  {383} (\bibinfo {year} {1967})}\BibitemShut {NoStop}%
\bibitem [{\citenamefont {Clarke}\ and\ \citenamefont
  {Braginski}(2004)}]{Clarke2004}%
  \BibitemOpen
  \bibfield  {author} {\bibinfo {author} {\bibfnamefont {J.}~\bibnamefont
  {Clarke}}\ and\ \bibinfo {author} {\bibfnamefont {A.~I.}\ \bibnamefont
  {Braginski}},\ }\href@noop {} {\emph {\bibinfo {title} {The SQUID
  handbook}}},\ Vol.~\bibinfo {volume} {1}\ (\bibinfo  {publisher}
  {Wiley-VCH},\ \bibinfo {year} {2004})\BibitemShut {NoStop}%
\end{thebibliography}

%
\end{document}